\documentclass[preprint,5p,authoryear]{elsarticle}
\usepackage{amssymb,amsmath}
\usepackage{graphicx}
\usepackage{dcolumn}
\usepackage{subfigure}
\usepackage{color}
\usepackage{soul}
\usepackage{float}
\usepackage{times}
\usepackage{soul}
\usepackage{xspace}
\usepackage[pdftex,breaklinks,colorlinks,
linkcolor=blue,
anchorcolor=blue,
citecolor=blue,
urlcolor=blue,]{hyperref}
\usepackage{lscape}

\usepackage[dvipsnames]{xcolor}
\usepackage{nameref}

\newcommand{\del}[1]{\!\xspace}

\usepackage[fulladjust]{marginnote}
\newcommand{\mpar}[1]{}
\newcommand{\mpardn}[1]{}

\newcommand{\new}[1]{{\color{black} #1}}
\newcommand{\mnote}[1]{}

\renewcommand{\eqref}[1]{\textup{{Eq.~(\ref{#1}})}}

\newcommand{\figref}[1]{\textup{{Figure~\ref{#1}}}}
\newcommand{\secref}[1]{\textup{{Section~\ref{#1}}}}
\makeatother

\makeatletter
\def\ps@pprintTitle{%
   \let\@oddhead\@empty
   \let\@evenhead\@empty
   \def\@oddfoot{\reset@font\hfil\thepage\hfil}
   \let\@evenfoot\@oddfoot
}
\makeatother

\begin{document}


\title{The impact of realistic axonal shape on axon diameter estimation using diffusion MRI}

\author[cbi,cai2r]{Hong-Hsi Lee\corref{cor}}
\ead{Honghsi.Lee@nyulangone.org}
\author[cfin,astro]{Sune N. Jespersen}
\author[cbi,cai2r]{Els Fieremans}
\author[cbi,cai2r]{Dmitry S. Novikov}

\cortext[cor]{Corresponding author}

\address[cbi]{Center for Biomedical Imaging, Department of Radiology, New York University School of Medicine, New York, New York, USA}
\address[cai2r]{Center for Advanced Imaging Innovation and Research (CAI2R), New York University School of Medicine, New York, New York, USA}
\address[cfin]{CFIN/MINDLab, Department of Clinical Medicine, Aarhus University, Aarhus, Denmark}
\address[astro]{Department of Physics and Astronomy, Aarhus University, Aarhus, Denmark}

\begin{keyword}
Monte Carlo simulation \sep diffusion MRI \sep axonal diameter mapping \sep caliber variation \sep axonal undulation \sep diffusion coarse-graining
\end{keyword}

\begin{abstract}
\noindent
To study axonal microstructure with diffusion MRI, axons are typically modeled as straight impermeable cylinders, whereby the transverse diffusion MRI signal can be made sensitive to the cylinder's inner diameter. 
However, the shape of a real axon varies along the axon direction, which couples the longitudinal and transverse diffusion of the overall axon direction. 
Here we develop a theory of the intra-axonal diffusion MRI signal based on coarse-graining of the axonal shape by 3-dimensional diffusion. We demonstrate how the estimate of the inner diameter is confounded by the  diameter variations (beading), and by the local variations in direction (undulations) along the axon.  We analytically relate diffusion MRI metrics, such as time-dependent radial diffusivity $D_\perp(t)$ and kurtosis $K_\perp(t)$, to the axonal shape, and validate our theory using Monte Carlo simulations in \new{synthetic undulating axons with randomly positioned beads,} and in realistic axons reconstructed from electron microscopy images of mouse brain white matter. We show that 
(i) \mnote{R2.1}\new{In the narrow pulse limit, the inner diameter from $D_\perp(t)$ is overestimated by about twofold due to a combination of axon caliber variations and undulations (each contributing a comparable effect size); 
(ii) The narrow-pulse kurtosis $K_\perp|_{t\to\infty}$ deviates from that in an ideal cylinder due to caliber variations; we also numerically calculate the fourth-order cumulant for an ideal cylinder in the wide pulse limit, which is relevant for inner diameter overestimation;
(iii) In the wide pulse limit, the axon diameter overestimation is mainly due to undulations at low diffusion weightings $b$; 
and (iv) The effect of undulations can be considerably reduced by directional averaging of high-$b$ signals, with the apparent inner diameter given by a combination of the axon caliber (dominated by the thickest axons), caliber variations, and the residual contribution of undulations.}  
\end{abstract}

\date{\today}



\maketitle


\section{Introduction and the summary of the results}
Diffusion MRI (dMRI) is the prime non-invasive method to evaluate microstructure of neuronal tissue, since the diffusion length at clinical and preclinical diffusion times is of the order of the cell size. Varying the diffusion time provides a gradual coarse-graining of the microstructure over the increasing diffusion length $L_d(t)$, revealing the corresponding structural details commensurate with $L_d(t)$ \citep{novikov2019note}. 

In brain white matter (WM), intra- and extra-axonal signals provide competing contributions, with the extra-axonal time dependence dominant at low (clinical) diffusion weightings \citep{burcaw2015meso,fieremans2016time,lee2018rd}. At sufficiently strong diffusion weightings ($b$-values), when the extra-axonal water signal is exponentially suppressed, the remaining intra-axonal signal can be used to quantify inner axonal diameters.
Indeed, the response of an ideal infinitely-thin fiber (``stick''), yielding the asymptotic $1/\sqrt{b}$ scaling of the directionally-averaged dMRI signal, was observed in vivo in human brain WM up to $b\leq$ 10 ms/$\mu$m\textsuperscript{2}  
\citep{mckinnon2017highb,veraart2019highb}, pointing at the lack of sensitivity to inner diameters on clinical scanners. 
The deviation from the $1/\sqrt{b}$ scaling, such as an extrapolated negative signal intercept at $1/\sqrt{b}\to 0$, was observed for 
$b\gtrsim$ 20 ms/$\mu$m\textsuperscript{2}, proving that the dMRI signal in this regime can reveal the elusive information about the axonal diameters or size \citep{veraart2020highb}. 
Inner axonal diameters can also be probed via diffusion-weighted spectroscopy of  metabolites \citep{ronen2014cc,palombo2016metabolite}. 
For the purposes of this so-called axonal diameter mapping (ADM), axons have been conventionally modeled as highly aligned perfect cylinders \citep{assaf2008axcaliber,Barazany2009,alexander2010activeax,duval2015spinal,benjamini2016wmadm,sepehrband2016adm}, with the effect of non-cylindrical axonal shapes 
considered only recently \new{\mpar{R2.5}\citep{palombo2018metabolite,lee2019axial}}. 

So far, ADM has incorporated the confounding effect of the variation between individual axons, resulting in the notion of effective MR radius \citep{burcaw2015meso}, dominated by the $4^{\rm th}$-order and  
$6^{\rm th}$-order moments of the cylinder radius distribution for narrow and wide pulse limits, respectively, in the practical case when there is not enough sensitivity to resolve the whole cylinder distribution \citep{assaf2008axcaliber,Barazany2009}. 
Furthermore, to factor out axonal orientation dispersion, dMRI signals with the same diffusion weighting $b$ can be averaged over the directions \citep{jespersen2013pfg,lasic2014magicangle,szczepankiewicz2015magicangle,kaden2016smt,veraart2019highb}.

\begin{figure}[t!]\centering
\includegraphics[width=0.475\textwidth]{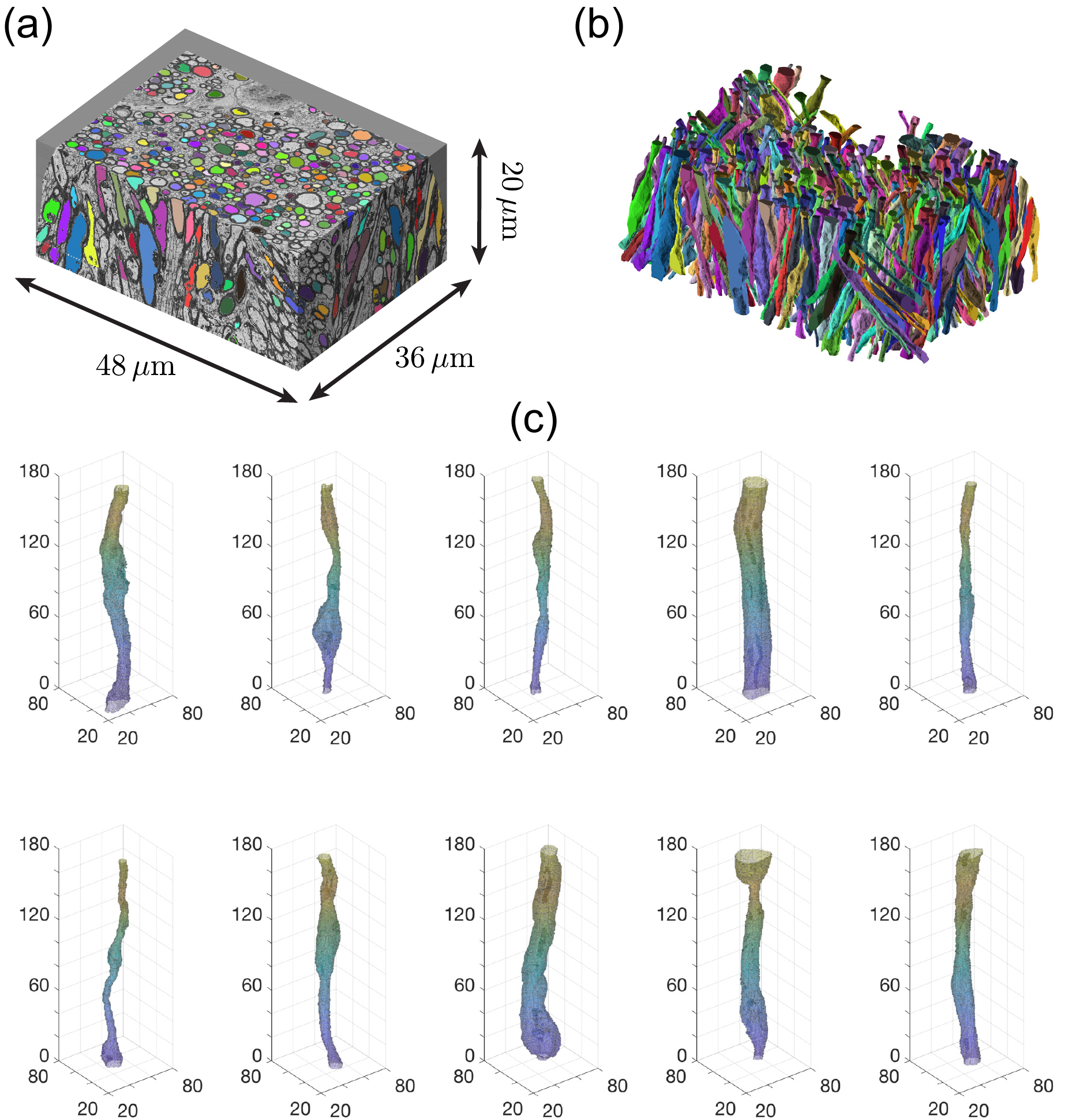}
\caption[]{(a) Realistic microstructure of the intra-axonal space reconstructed from 3{\it d} SEM images of the mouse brain genu of corpus callosum, and (b) its 3{\it d} representation. (c) Segmented IASs were aligned along the $z$-axis to control the orientation dispersion. Only long axons were chosen and cropped into 18 $\mu$m in length. \new{The mean diameter is $\sim 1$ $\mu$m.} The units along $x$-, $y$-, and $z$-axis are 0.1 $\mu$m. (Adapted from \citep{lee20193dem} with permission from Springer.)
}
\label{fig:IAS-shape}
\end{figure}

\begin{figure}[th!!]\centering
\includegraphics[width=0.475\textwidth]{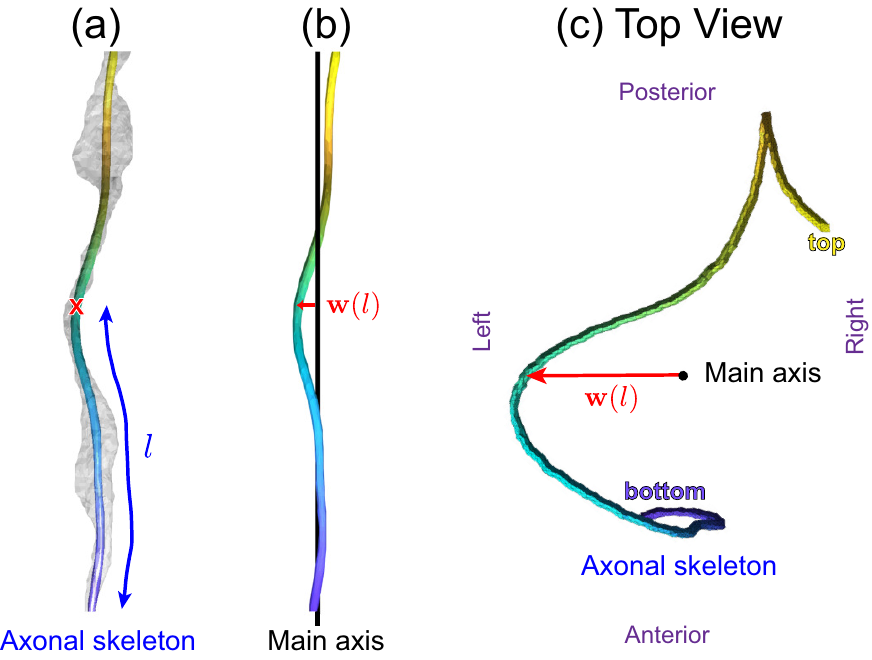}
\caption[]{Axonal undulation.  \mnote{R1.5\\R2.14}
(a) The axonal skeleton is a curve connecting the center of mass of each cross-section along an axon. For a given point on the skeleton (red cross mark), $l$ is the axonal length accumulated from an end of the skeleton to the given point. (b) The deviation ${\bf w}(l)$ is the shortest distance from the skeleton to the main axis. (c) The top view of the axonal skeleton, the main axis, and the deviation ${\bf w}(l)$ in (b).}
\label{fig:appendix-undulation-shape}
\end{figure}

In this work, we consider \mpar{R1.4}\new{ following confounding factors of ADM stemming} from realistic, non-cylindrical axonal shapes (\figref{fig:IAS-shape}): {\it axonal caliber variations} \mpar{R2.2}(changes in the axonal cross-section, also referred to as beading or swelling), and {\it undulations} (changes of local directionality relative to the overall axon direction, \figref{fig:appendix-undulation-shape}), that have been observed in histological studies in WM \citep{shepherd2002varicosity,shepherd2003varicosity,li2008reperfusion,budde2010bead,tang2012microtubule,baron2015stroke,bain2000undulation,bain2003undulation,ronen2014cc,schilling20163dconfocal,schilling2018histmri,lee20193dem,abdollahzadeh20193dem}.

\mpar{R1.4}\new{In previous studies, simulations targeting undulations} were performed in artificially designed undulating thin fibers, showing that the inner diameters of highly undulating fibers are significantly overestimated \new{\citep{nilsson2012undulation,brabec2019undulation}}\mpar{R2.5}. However, the influence of both undulations and axon caliber variations on diffusion metrics and diameter estimations has not yet been investigated for realistic axonal shapes.

Technically, the problem of non-cylindrical irregular axons becomes 3-dimensional and involves an interplay between the diffusion diffraction due to confinement in the two transverse dimensions, and an unconfined longitudinal diffusion. Here we show that the coarse-graining of transverse axonal cross-sections over the diffusion length along the axon results in a ``coherent averaging" of such two-dimensional signals. The resulting  transverse signal cumulants acquire sensitivity to axonal caliber variations \citep{budde2010bead,baron2015stroke}, and  undulations \citep{nilsson2012undulation,brabec2019undulation}.

\begin{figure}[t!]\centering
\includegraphics[width=0.32\textwidth]{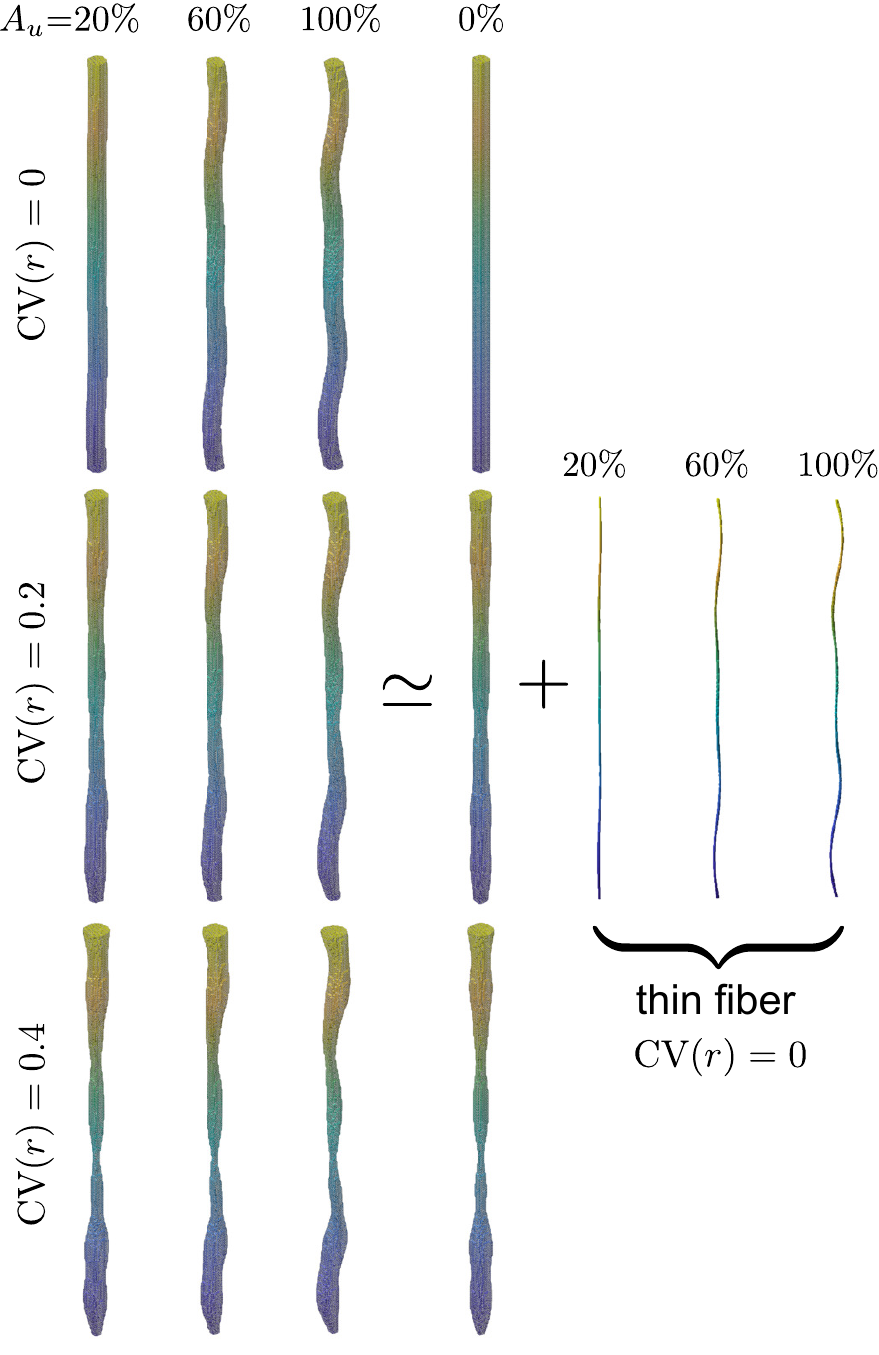}
\caption[]{\reversemarginpar\mnote{R1.1\\R2.3}3$d$ geometries of \new{synthetic axons with randomly positioned beads along $z$-axis and sinusoidal undulations along $x$- and $y$-axes. 
The strength of caliber variation is tuned based on the coefficient of variation of the radius $\text{CV}(r)=0-0.4$, and the amplitude of undulation is scaled via a factor $A_u=0-100$\%.
In the narrow pulse limit, diffusivity transverse to axons is well approximated by a sum of the independent contributions from pure caliber variations (no undulation) and pure undulations (no caliber variation).}
}
\label{fig:bead-shape}
\end{figure}

In what follows, we develop an analytical theory for the time-dependent diffusion coefficient and kurtosis due to beading and undulations, and validate it using 3-dimensional (3{\it d}) Monte Carlo (MC) simulations in artificially designed \new{undulating axons with randomly positioned beads} (\figref{fig:bead-shape}), as well as in realistic microstructure of intra-axonal space (IAS) segmented from 3{\it d} scanning electron microscopy (SEM) images of mouse brain corpus callosum,  \figref{fig:IAS-shape} \citep{lee20193dem}. 
Combining theory with MC simulations, here we show: 
\new{
\begin{enumerate}[(i)]

\item In the narrow pulse limit, we find the signal's cumulants (time-dependent diffusivity and kurtosis) due to caliber variations, coarse-grained over the axon length (Sec.~\ref{sec:coh}), and the approximate ansatz for the full 3-dimensional diffusion propagator in the long-time limit, Eq.~(\ref{eq:G-ansatz}) in \ref{sec:app-averaging}. 
We also find the cumulants separately for thin undulating axons (Sec.~\ref{sec:theory-undulation}). 
It turns out that caliber variations and undulations provide roughly equal contributions to the overall $D_\perp(t)$, that together lead to an overestimation of the axonal diameter by about twofold relative to a typical diameter within a bundle, Figs. \ref{fig:bead-result} and \ref{fig:IAS-result}. 

\item The fourth-order cumulant term, $\sim b^2$:  
For narrow pulses, the radial intra-axonal kurtosis $K_\perp(t)|_{t\to\infty}$ becomes significantly different from the  value $-1/2$ for a perfectly straight cylinder \mpar{R2.4} \citep{burcaw2015meso} due to caliber variations, Sec.~\ref{sec:coh}, Figures \ref{fig:bead-result} and \ref{fig:IAS-result}. 
For wide pulses, we calculate for the first time this term numerically for a cylinder, Sec.~\ref{sec:caliber-wide}, and discuss its effect on diameter overestimation when applying strong gradients. 

\item In the wide pulse limit, when the pulse width $\delta \sim \Delta$ is of the order of gradient separation $\Delta$, the overestimation of axonal diameter in the low-$b$ regime, based on $D_\perp(\Delta,\delta)$, is mainly due to undulations (Sec.~\ref{sec:und-wide}), which completely overshadows the ``diameter'' mapping for both synthetic and realistic axons, Figures \ref{fig:lissajous-wide} and \ref{fig:IAS-wide}. 

\item Finally, we quantitatively analyze the way to minimize the undulation effect on axon diameter estimation by directional average of the signal at high $b$, Sec.~\ref{sec:dir-avg}, and study the corresponding ADM results for different sequence timings, \figref{fig:IAS-highb}. 

\end{enumerate}
The conclusions made for individual axons are generalized to a collection of multiple axons, as the net signal (and signal metrics such as diffusivities) are found by taking the volume-weighted sum of 227 individual axon contributions.\mpar{R4.1}
}


\section{Theory} 
\label{sec:theory}

Here we consider the theory of caliber variations and undulation effects first in the narrow pulse limit (Sec.~\ref{sec:NP}), and then for more realistic wide pulses (Sec.~\ref{sec:WP}), followed by the effects of directional averaging (Sec.~\ref{sec:dir-avg}). 
The theory necessary for the result (i) formulated above is obtained in Sections \ref{sec:coh} and \ref{sec:theory-undulation}; 
for the result (ii), in Sec.~\ref{sec:coh} and \ref{sec:caliber-wide}; 
for the result (iii), in Sec.~\ref{sec:WP}; 
and for the result (iv), in Sec.~\ref{sec:dir-avg}.

\subsection{Narrow pulse limit}
\label{sec:NP}

\subsubsection{Caliber variations: Coherent averaging} 
\label{sec:coh}

Henceforth, we define caliber variation as a varying cross-sectional area along an axon. 
In ADM from dMRI, the diameter or radius is used to evaluate the axonal size. 
In this study, we use the equivalent circle diameter/radius (axon caliber) calculated based on the cross-sectional area \citep{west2016gratio}. \mpar{R2.6}\new{(Alternatively, fitting an ellipse to an axon cross-section also provides size estimations based on short and long axis lengths, whose geometric mean is close to the equivalent circle diameter since both metrics are related to the cross-sectional area \citep{lee20193dem}; we will not consider these metrics here.)}

Our starting point is the diffusion-diffraction long-time narrow-pulse limit of the diffusion signal for a 2{\it d} confined pore \citep{callaghan1991pore,callaghan1991book}, cf. \ref{sec:app-averaging}: 
\begin{equation}\label{eq:S}
S({\bf q}_\perp,t\to\infty)\simeq \frac{1}{A^2}|\rho({\bf q}_\perp)|^2 \, ,
\end{equation}
where the non-diffusion weighted signal is normalized to $S|_{q_\perp=0}\equiv 1$, ${\bf q}_\perp$ is the diffusion wave vector within the plane of the pore, $t$ is the diffusion time, $A$ is the area of the 2{\it d} pore, 
and $\rho({\bf q}_\perp)$ is the pore form factor, i.e., the Fourier transform of the pore shape
\begin{equation} \label{eq:poreshape}
\rho({\bf x})=\left\{\begin{array}{l}1,\quad\text{inside\,pore,}\\0,\quad\text{outside\,pore.}\end{array}\right.
\end{equation}
When diffusion can occur in the third dimension, this equation is modified, as discussed in \ref{sec:app-averaging}:
\begin{equation}\label{eq:S-cg}
S({\bf q}_\perp,t\to\infty)\simeq \frac{1}{A^2} \left|\langle \rho({\bf q}_\perp) \rangle_{L_d}\right|^2 \,,
\end{equation}
where $\langle...\rangle_{L_d}$ is the {\it coherent averaging} 
of the form factor $\rho({\bf q}_\perp)$ of the two-dimensional cross-sections along the axonal skeleton, and $A$ is the mean cross-sectional area. 
Technically, this averaging is performed within a \mpar{R2.8}\new{window on the scale of the diffusion length} $L_d(t)\sim \sqrt{2D_a t}$ along the axon, 
with the effective intra-axonal diffusivity $D_a(t)|_{t\to \infty}$, according to \eqref{eq:S-qperp-t} in \ref{sec:app-averaging}. 
Practically, for sufficiently long $L_d(t)\gg l_c$ exceeding the correlation length  $l_c$ of the placements of axonal beadings or swellings, the average in \eqref{eq:S-cg} can be extended over the length of each axon, as in \eqref{eq:segment}. 

We refer to the averaging in \eqref{eq:S-cg} as coherent because it is performed before the absolute value is taken (i.e., accounting for the phase information in the form factor). 
Adding the contributions from different axons would amount to a further volume-weighted {\it incoherent averaging} of \eqref{eq:S-cg} over all axon form factors (or over the distant parts of the same axons, separated by much longer than $L_d(t)$). 
Therefore, we average coherently over the spins within each connected compartment (axon) (more generally, within domains $\sim L_d(t)$ that they are able to explore), and then incoherently over the non-exchanging compartments (\secref{sec:theory-incoherent}). 

For example, for a cylinder of varying radii without undulations (e.g., \figref{fig:bead-shape}), the form factor for a given cross-section of radius $r = r(z)$, 
\begin{equation*} 
\frac{\rho(q_\perp)}{\pi r^2}=\int_{|{\bf x}_\perp|<r}\! \frac{d^2 {\bf x}_\perp}{\pi r^2} \, e^{i{\bf q}_\perp\cdot{\bf x}_\perp} = \frac{2J_1(q_\perp r)}{q_\perp r} \,,
\end{equation*} 
with \mpar{R2.10\\R2.12}\new{${\bf x}_\perp=(x_1,x_2)$ the spatial coordinate transverse to the main axis ${\bf \hat{z}}$, } and $J_\nu(\cdot)$ the Bessel function of the first kind, needs to be coherently averaged over all cross-sections. 
We can do it term-by-term in its Taylor expansion,\mpar{R2.11}
\begin{equation*} 
\frac{\rho(q_\perp)}{\pi r^2} = 1-\tfrac{1}{8}(q_\perp r)^2 \new{+}\tfrac{1}{192}(q_\perp r)^4+...\,,
\end{equation*} 
such that 
\begin{equation}\label{eq:V-cylinder}
\frac{1}{A} \langle \rho(q_\perp)\rangle_{L_d} = 1-\tfrac{1}{8}q_\perp^2\langle r^2\rangle_v + \tfrac{1}{192}q_\perp^4\langle r^4\rangle_v + ... \,,
\end{equation}
where $A=\pi\langle r^2\rangle$, and $\langle...\rangle_v$ denotes the volume-weighted (i.e., area-weighted) quantity for the disk with area $\propto r^2$, i.e., 
$$\langle r^n\rangle_v \equiv \langle r^{n+2}\rangle/\langle r^2\rangle. $$

Substituting \eqref{eq:V-cylinder} into \eqref{eq:S-cg} and comparing with the moments expansion of the signal,
\begin{equation*}
S(q_\perp,t) \simeq 1-\tfrac{1}{2!}q_\perp^2 \langle \delta x^2\rangle + \tfrac{1}{4!}q_\perp^4 \langle \delta x^4 \rangle + ...\,,
\end{equation*}
we obtain the moments of the diffusion displacement $\delta x$ along a direction perpendicular to the axon:
\begin{subequations} \label{eq:x2-x4-r} \begin{align} \label{eq:x2-r}
\langle \delta x^2 \rangle &= \tfrac{1}{2}\langle r^2\rangle_v\,,\\ \label{eq:x4-r}
\langle \delta x^4 \rangle &= \tfrac{1}{4}\langle r^4\rangle_v + \tfrac{3}{8}\langle r^2\rangle_v^2\,,
\end{align} \end{subequations}
which yield the diffusivity and kurtosis transverse to axons based on definitions \citep{jensen2005dki,jensen2010dki}
\begin{subequations} \label{eq:D-K-def} \begin{align} \label{eq:D-def}
D_\perp(t) &\equiv \frac{\langle \delta x^2\rangle}{2t}\,,\quad\\ \label{eq:K-def}
K_\perp(t) &\equiv \frac{\langle \delta x^4\rangle}{\langle \delta x^2\rangle^2}-3\,.
\end{align} \end{subequations} 
%
In particular, substituting \eqref{eq:x2-r} into \eqref{eq:D-def} yields the radial diffusivity (RD)
\begin{equation} \label{eq:reff}
D_\perp(t)=\frac{\langle r^2\rangle_v}{4t}\,,\quad
\langle r^2\rangle_v\equiv\frac{\langle r^4\rangle}{\langle r^2\rangle}\,,
\end{equation}
generalizing the concept of dMRI-sensitive radius \citep{burcaw2015meso} for variable axonal shape, 
$\langle r^2\rangle_v$.
We emphasize that \eqref{eq:reff} is only applicable to an axon with no undulations. 

Similarly, substituting \eqref{eq:x2-x4-r} into \eqref{eq:K-def} yields the radial kurtosis (RK) in the $t\to\infty$ limit
\begin{equation}\label{eq:K}
K_\infty=\frac{\langle r^4\rangle_v}{\langle r^2\rangle_v^2}-\frac{3}{2}=\frac{\langle r^6\rangle \langle r^2\rangle}{\langle r^4\rangle^2}-\frac{3}{2}\,,
\end{equation}
whose minimum value
is $-1/2$ for a perfectly straight cylinder with a constant radius \mpar{R2.4}\new{\citep{burcaw2015meso}}, as follows from the Cauchy-Schwarz inequality, indicating that the microstructural heterogeneity along axons due to caliber variation causes the intra-axonal kurtosis to increase as compared to the case of perfectly straight axons with no caliber variation. 

\subsubsection{Incoherent averaging in non-exchanging compartments} \label{sec:theory-incoherent}
Incoherent averaging over non-exchanging compartments, such as multiple axons and the extra-axonal space, is performed 
for the {\it signals}, according to their $T_2$-weighted fractions $f_i$: 
\begin{equation} \label{eq:incoh}
S({\bf q}_\perp, t) = \sum_{i=1}^N f_i S_i({\bf q}_\perp, t) + (1-f) S_e({\bf q}_\perp, t) \,, \quad \sum_{i=1}^N f_i=f 
\end{equation}
for a fiber bundle of $N$ axons with the intra- and extra-axonal fractions ($f$ and $1-f$) \new{and signals ($S_i$ and $S_e$)}, respectively. 
Expanding \eqref{eq:incoh} into diffusion cumulants, the overall diffusivity and kurtosis are then given by \citep{jensen2005dki,dhital2018restricted}
\begin{subequations} \label{eq:D-K-intra-extra} \begin{align}
\overline{D}&=\sum_i f_i\cdot D_i + (1-f)\cdot D_e\,,\\
\overline{K}&=3\frac{\text{var}(D)}{\overline{D}^2} + \frac{\langle D^2K\rangle}{\overline{D}^2}\,,
\end{align} \end{subequations} 
where
\begin{subequations} \begin{align*}
\text{var}(D)&=\sum_i f_i\cdot (D_i-\overline{D})^2 + (1-f)\cdot(D_e-\overline{D})^2\,,\\
\langle D^2K\rangle &= \sum_i f_i\cdot D_i^2 K_i + (1-f)\cdot D_e^2 K_e
\end{align*} \end{subequations}
in terms of intra-axonal diffusivities and kurtoses $D_i$ and $K_i$, and extra-axonal ones $D_e$ and $K_e$, respectively. 

In this study, we will focus on the coherent averaging for an individual axon, and discuss the effect of the incoherent averaging over all axons and the extra-axonal space in \secref{sec:discuss-inter-time}.

\subsubsection{Axonal undulations, narrow pulse limit}
\label{sec:theory-undulation}

Axonal undulations (\figref{fig:appendix-undulation-shape}) can be parameterized by the 2-dimensional deviation ${\bf w}(l)$ of the axonal skeleton from the axon's main axis, where ${\bf w}(l)$ is the shortest distance from the skeleton to the main axis, and $l$ is a 1{\it d} coordinate along the skeleton, defined as the axonal length accumulated from one end of the skeleton to a given point on the skeleton. 

\new{In our EM sample, segmented axons are generally aligned along the $\bf\hat{z}$ direction, and the axonal skeleton is calculated by simply connecting the center of mass of the cross-section for each slice. The main axis is a line crossing the skeleton center (mean of spatial coordinates of the skeleton) and aligned along the mean direction of the skeleton (parallel to the line connecting two ends of the skeleton). For axons aligned roughly perpendicular to the $\bf\hat{z}$ direction, the skeleton can be calculated by applying a distance transform within the segmentation and searching for the local maxima in the distance map.}\mpar{R2.9}

First, let us study the effect of pure undulation, assuming that the inner diameter is so small that diffusion displacement perpendicular to the axonal skeleton is negligible. In this case, only the diffusion displacement along the axonal skeleton contributes to the intra-axonal diffusivity, and the diffusion along the skeleton can be treated as a 1{\it d} {\it Gaussian diffusion}; the nontrivial physics arises from the projection of the Gaussian propagator along the skeleton onto the transverse plane.

Considering the diffusion along an undulated skeleton ${\bf l} = {\bf w} + z{\bf\hat{z}}$, the second-order cumulant of the diffusion displacement projected onto the direction $\bf\hat{n}$ is given by \\\citep{bouchaud1990review}
\begin{equation} \label{eq:l-n-2}
    \langle (\delta{\bf l}\cdot{\bf\hat{n}})^2\rangle \simeq \frac{1}{C} \int \left| ({\bf l} - {\bf l'})\cdot{\bf\hat{n}} \right|^2\cdot e^{-\frac{(l-l')^2}{4D_a t}}\, dl\, dl'\,,
\end{equation}
where the integral extends over the length $L$ of the skeleton, $D_a$ is the diffusivity along the skeleton, and $C$ is a normalization factor:
\begin{equation*}
C=\int e^{-\frac{(l-l')^2}{4D_at}}\,dl\,dl' = \sqrt{4\pi D_a t} \cdot L\,.
\end{equation*}
Then the second-order cumulant $\langle \delta w^2 \rangle$ of the deviation perpendicular to the main axis, i.e., $\langle \delta w_x^2\rangle + \langle \delta w_y^2\rangle$ summed over $\bf\hat{x}$ and $\bf\hat{y}$ directions, is given by 
\begin{equation} \label{eq:w2}
\langle \delta w^2(t)\rangle \simeq \frac{1}{C}\int \left|{\bf w}(l)-{\bf w}(l')\right|^2 \cdot e^{-\frac{(l-l')^2}{4D_a t}} \, dl \, dl'\,.
\end{equation}
The nonzero time-dependent $\langle \delta w^2(t)\rangle$ results in a contribution $\langle \delta w^2\rangle/(2d\cdot t)$ to the RD in {\it d} = 2 dimension.
Similar considerations were recently put forward by  \cite{brabec2019undulation}.

Let us now consider the combined radial diffusion: coming from the diffusion confined within the finite-caliber axonal cross-section, \eqref{eq:reff}), as well as that along the undulating axonal skeleton and projected onto the transverse direction. Assuming statistical independence of the caliber variations and undulations, justifiable due to the fact that undulations \new{($\sim$20-30 $\mu$m in wavelength \citep{fontana1781undulation})}\mpar{R1.2} typically occur on a much larger scale than the correlation length of caliber variation \new{($\sim$1 $\mu$m, similar to  axon diameter \citep{lee20193dem})}, leads to the statistical independence of the transverse displacements due to caliber variations and undulations, such that the variances add up, and 
the RD \new{in the narrow pulse limit} is given by
\begin{subequations} \label{eq:Dt-narrow}
\begin{align} \label{eq:Dt}
D_\perp(t) & \equiv \frac{r_\text{eff}^2}{4t}\\ \label{eq:D-caliber-undulation}
&\simeq \frac{\langle r^2\rangle_v}{4t} + \frac{\langle \delta w^2\rangle_{t\to\infty}}{4t}\,,
\end{align}
\end{subequations}
where $\langle \delta w^2\rangle_{t\to\infty}$ is calculated based on \eqref{eq:w2} at long diffusion times. The discrepancy between the dMRI-sensitive effective radius $r_\text{eff}^2$ and the $2d$  histology-based prediction including only caliber variation $\langle r^2\rangle_v$ is then given by:
\begin{equation} \label{eq:dreff}
r_\text{eff}^2-\langle r^2\rangle_v \simeq \langle \delta w^2\rangle_{t\to\infty}\,,
\end{equation}
indicating that the axonal size can be overestimated due to the axonal undulation.

\new{Furthermore, for a single length scale in microgeometry, such as the bead distance or undulation wavelength, the RK scales as $1/t$ in the narrow pulse limit (see Eq. [53] in \citep{novikov2010emt}):
\begin{equation} \label{eq:Kt}
    K_\perp(t) \simeq K_\infty + \frac{c_K}{t}\,,
\end{equation}
with $K_\infty$ in \eqref{eq:K} and the kurtosis time-dependence amplitude $c_K$. Here we ignore the undulation effects on the theory of $K_\infty$ for the simplicity. 
}

\new{
\subsection{Wide pulse limit} 
\label{sec:WP}

In sections \ref{sec:coh} and \ref{sec:theory-undulation}, we focused on diffusion transverse to axons in the narrow pulse limit, i.e., diffusion gradient pulse width $\delta\ll t_D$, where
\begin{equation} \notag
t_D = r^2/D_0
\end{equation}
is the  time for a spin to diffuse across an axon, and $D_0=D_a(t)|_{t\to 0}$ is the intrinsic axoplasmic diffusivity.
In the opposite limit, i.e., the wide pulse limit $\delta\gg t_D$,  the RD has different functional forms from \eqref{eq:Dt-narrow}. It is less trivial to disentangle the effects of caliber variations and undulations in this limit. Instead, we provide solutions when one of the contributions dominates.
\mpar{R1.3\\R4.2}

\subsubsection{Sensitivity to axonal caliber, wide pulse limit}
\label{sec:caliber-wide}

In the wide pulse limit, the signal attenuation transverse to a straight axon with no caliber variations is given by \citep{neuman1974gpa,lee2018rd}
\begin{equation} \label{eq:S-wide-pulse}
\ln S = -c_1\cdot\frac{g^2r^4}{D_0}\delta - c_2\cdot\frac{g^4r^{10}}{D_0^3}\delta + {\cal O}(g^6)\,,
\end{equation}
where Larmor frequency gradient $g\equiv\gamma G$ is defined by the  gyromagnetic ratio $\gamma$ and the diffusion gradient $G$. For reference, $g=$ 0.0107 ($\mu$m$\cdot$ms)$^{-1}$ for $G=$ 40 mT/m. The constants $c_1$ and $c_2$ are related to the shape of axon cross-section. 
For a circular cylinder cross-section, 
\begin{equation} \label{c1c2}
    c_1=\frac7{48}\approx 0.1458 \,, \quad c_2\approx0.0022 \,.
\end{equation}
\new{
The constant $c_1$ was analytically derived by \citet{neuman1974gpa}; 
the constant $c_2$ is here for the first time numerically estimated via MC simulations in \ref{sec:app-cyl}.} 

We now define the transverse diffusivity in a standard way,
\begin{equation} \label{Dperp}
D_\perp(t,\delta) \equiv \left.-\tfrac{1}{b}\ln S\right|_{b\to0} 
\end{equation}
in the wide pulse limit, by denoting $t=\Delta$ (the distance between the fronts of the pulses). 
Here the diffusion weighting is given by
\begin{equation} \label{eq:b-value}
    b=g^2\delta^2(t-\delta/3)\,.
\end{equation}
Sending $b\to0$ above corresponds to $g\to0$.

Based on \eqref{eq:S-wide-pulse} and \eqref{Dperp}, the corresponding wide-pulse (``WP") transverse diffusivity due to caliber variations 
\begin{equation} \label{eq:D-neuman}
D_\text{cal, WP}(t,\delta) \simeq \left( \frac{c_1 \langle r^4\rangle_v}{D_0\delta}\right)\cdot\frac{1}{t-\delta/3}\,,
\end{equation}
where the idea of coarse-graining along the axon (resulting in volume-weighted averaging over adjacent cross-sections) is applied, assuming the long-time limit for the longitudinal direction, $\sqrt{2D_0 \delta}\gg l_c$, similar to Sec.~\ref{sec:coh} above. We will expand on the physical intuition at the end of this subsection. 

For any axon, we can define the effective radius 
\begin{equation} \label{eq:reff-wide}
    r_\text{eff, WP} \equiv \left(\frac{1}{c_1}\delta\cdot (t-\delta/3)\cdot D_0 D_\perp(t,\delta)\right)^{1/4}
\end{equation}
as an estimate of the axon size from $D_\perp(t,\delta)$ using a wide pulse dMRI sequence. 
From this definition, neglecting the undulations, and based on \eqref{eq:D-neuman}, we obtain
\begin{equation} \label{r62}
r_\text{eff, WP} \simeq r_\text{cal} \equiv \left(\langle r^4\rangle_v\right)^{1/4}\,,\quad \langle r^4\rangle_v \equiv \frac{\langle r^6 \rangle}{\langle r^2 \rangle}\,.
\end{equation}
Note the formal equivalence of \eqref{r62} and the first term of Eq.~(39) in \citep{burcaw2015meso}:
The effective radius for a {\it distribution} of ideal cylinders of different radii is given by the same ratio of the moments of the distribution as that for a single axon with a varying radius. 
The physical origin of this effect is the area-weighted $r^4$-dependence of the $g^2$ term in \eqref{eq:S-wide-pulse}.  \eqref{r62} is applicable to the {\it caliber regime}, when the effects of undulations can be either neglected 
(cf. \eqref{eq:reff-wide-undul}, which is so far not supported by our EM data (as we will see below), or factored out via directional averaging, Sec.~\ref{sec:dir-avg}. 
}


We now provide the physical intuition behind the above results in the case of arbitrary caliber variations, noting that the precise values of coefficients such as $c_1$ and $c_2$ are geometry-dependent. However, the dependence of \eqref{r62} on the 6th and 2nd order moments of the caliber variation distribution within an axon is quite general. 
%
We first recall \citep{lee2018rd,Kiselev2018} that the wide-pulse signal attenuation \eqref{eq:S-wide-pulse} 
within a single $2d$ disk can be viewed as a transverse relaxation process during the duration $2\delta$ of the gradient, $\ln S \sim - R_2^*(r) \cdot 2\delta$ 
with the rate constant $R_2^*(r) \sim g^2 r^4/D_0$, and the $g^4$ corrections to $R_2^*(r)$ appearing as an expansion in 
the typical accumulated phase $(g r)\cdot t_D$ on a single traverse of axonal cross-section. 

Next, the coarse-graining intuition allows us to map the 3{\it d} signal attenuation problem onto the transverse relaxation in a number of 2{\it d} ``compartments" with the relaxation rate constants  $R_2^*(r_i)$ corresponding to different axonal cross-sections $r_i(z)$ within the axonal segment of the order of the diffusion length along the axon, such that  $|z-z'| \lesssim L_d(\delta)$. 

The time scale to exchange between sufficiently distinct 2{\it d} ``compartments" is $t_c=l_c^2/D_0\simeq 10-20$ ms, with $l_c$ the correlation length along an axon (e.g., distance between swellings $= 5-6$ $\mu$m \citep{lee2019axial}) and $D_0 = 2$ $\mu$m$^2$/ms. In contrast, the relaxation time scale for each 2{\it d} compartment is quite long: $1/R_2^*(r_i)\sim D_0/(g^2r_i^4)\simeq 300$ ms even for $g = 0.08$ ($\mu$m$\cdot$ms)$^{-1}$ on Connectom scanners ($G= 300$ mT/m) and $r_i\sim 1$ $\mu$m.
The separation of scales $t_c\ll 1/R_2^*(r_i)$ allows us to consider those ``compartments'' in the fast exchange regime; in this case, the net transverse relaxation to the lowest order in $\delta$ will have a rate given by a weighted average, $R_2^* = \sum f_i R_2^*(r_i)$ with $f_i \propto r_i^2$ \new{\citep{zimmerman1957R2}}. 
This means that the effective radius in the wide-pulse regime 
due to the $r^4$ weighting in $g^2$ term in \eqref{eq:S-wide-pulse} will be of the same nature as that coming from different ideal cylinders with different radii \citep{burcaw2015meso}. 
\new{The $r^4$ weighting in diffusivity of wide pulse limit is generalized to $\langle r^4\rangle_v$ in \eqref{eq:D-neuman} for the same reason. However, for the higher order $g^4$ term, the effect of the time scale $t_c$ may be non-trivial; fortunately, as we will show below, the $g^4$ terms are practically not important.}

\new{
\subsubsection{Undulations, wide pulse limit}
\label{sec:und-wide}

For RD due to undulations in the wide pulse limit, we focus on a simplified 1-harmonic model. As derived in \ref{sec:app-undulation}, the pulsed-gradient measured RD due to 1-harmonic undulation is given by
\begin{multline} \label{eq:D-undul-wide-1h}
    D_u(t,\delta)
    \simeq \frac{1}{4}\cdot \frac{w_{0}^2 t_u^2}{\delta^2(t-\delta/3)} \cdot \left[ 2\frac{\delta}{t_u} -2 \right.\\ \left.+ 2e^{-t/t_u} + 2e^{-\delta/t_u} - e^{-(t-\delta)/t_u} - e^{-(t+\delta)/t_u} \right]\,,
\end{multline}
where $w_0$ is the undulation amplitude, and $t_u=\lambda^2/(4\pi^2D_a)$ is the correlation time corresponding to the undulation wavelength $\lambda$. In the wide pulse limit of undulations, i.e., $\delta\gg t_u$, the RD due to undulations becomes
\begin{equation} \label{eq:D-undul-neuman-1h}
    D_u(t,\delta)\simeq \frac{1}{8\pi^2}\cdot\frac{w_{0}^2\lambda^2}{D_a\delta}\cdot\frac{1}{t-\delta/3}\,, \quad \delta\gg t_u\,,
\end{equation}
which has exactly the same functional form of $t$ and $\delta$ as \eqref{eq:D-neuman}.
Using this similarity in functional forms, we here define an effective radius $r_\text{und}$ due to undulations by considering the {\it undulation regime}: When the length scale of axon caliber (diameter) is much smaller than that of undulation ($\sqrt{w_0\lambda}$), the RD is mainly contributed by undulations, i.e., $D_\perp\simeq D_u$. Substituting into \eqref{eq:reff-wide} with $D_a\simeq D_0$, we define $r_\text{und}$ due to undulations as an estimate of the ``axon size'' by using wide pulse dMRI sequence:
\begin{equation} \label{eq:reff-wide-undul}
    r_\text{eff, WP} \simeq r_\text{und} \equiv \left(\frac{1}{8\pi^2 c_1}\right)^{1/4} \sqrt{w_0\lambda} \approx 0.543\, \sqrt{w_0 \lambda}
\end{equation}
proportional to the geometric mean of undulation amplitude and wavelength.

To determine whether the simulated RD falls into the {\it caliber regime} or {\it undulation regime}, we calculate the effective radius $r_\text{eff, WP}$ of wide pulse sequence based on \eqref{eq:reff-wide} and simulated $D_\perp$, and compare with the predictions of caliber regime, $r_\text{cal}$ in \eqref{r62}, and of undulation regime, $r_\text{und}$ in \eqref{eq:reff-wide-undul}, cf. Figures \ref{fig:lissajous-wide} and \ref{fig:IAS-wide}.
}

\new{
\subsection{Directionally averaged signal decay} \mpar{R3.1}
\label{sec:dir-avg}

Rotationally invariant axon size estimation is based on the directional average of the signals for each b-shell, and on the  analysis of the deviations from the $1/\sqrt{b}$ scaling \citep{veraart2019highb,veraart2020highb}:
\begin{equation} \label{eq:S-sm}
    \overline{S}(b)\simeq\beta e^{-bD_\perp+{\cal O}(b^2)} b^{-1/2}\,,
\end{equation}
where $\beta=\sqrt{\pi/(4D_a)}$ (we set axonal water fraction to unity since we are studying only the intra-axonal signal), and the estimated RD, $D_\perp$, yields a axon size estimation based on Eqs.~(\ref{eq:Dt}) and (\ref{eq:reff-wide}) in narrow and wide pulse limit, respectively.

Averaging diffusion signals isotropically over each b-shell has been shown to factor out axonal orientation dispersion \citep{jespersen2013pfg,kaden2016smt}. Furthermore, at short times $t,\delta\ll\lambda^2/D_a$, water molecules diffuse only within a short axon segment ($\sim L_d(t)\ll \lambda$) with limited undulations, and thus the directional average can factor out the effect of undulations on RD and on the axon size estimation.
}

\section{Methods}
All procedures performed in studies involving animals were in accordance with the ethical standards of New York University School of Medicine. All mice were treated in strict accordance with guidelines outlined in the National Institute of Health Guide for the Care and Use of Laboratory Animals, and the experimental procedures were performed in accordance with the Institutional Animal Care and Use Committee at the New York University School of Medicine. This article does not contain any studies with human participants performed by any of the authors.

\new{
\subsection{Artificial substrates for simulations: \\ Undulating axons with randomly positioned beads} \label{sec:syn-fiber}
}
\mpar{R1.1}\new{To test the applicability of Eqs.~(\ref{eq:reff}) and (\ref{eq:K}) and their generalized forms (\ref{eq:Dt-narrow}) and (\ref{eq:Kt}) for narrow pulses and \eqref{eq:D-neuman} for wide pulse, we designed undulating axons with randomly positioned beads (\figref{fig:bead-shape}) with pore shape
\begin{equation} \label{rho-gen}
    \rho({\bf x}) = \left[{\bf w}(l(z))+z{\bf \hat{z}}\right] * v(r(z))\,,
\end{equation}
where $v(r)$ is a sphere of radius $r$, the operator $*$ is the convolution, and ${\bf w}(l)$ is the deviation of skeleton from the main axis, generated by using the Lissajous curve
\begin{equation} \label{eq:lissajous}
    {\bf w}(l(z)) = A_u\cdot\left[ w_x \sin(k_x z + \phi_x) {\bf \hat{x}} + w_y \sin(k_y z) {\bf \hat{y}} \right]\,,
\end{equation}
with amplitudes $(w_x,w_y)=(0.5,0.7)$ $\mu$m, wave vectors $k_{x,y}=2\pi/\lambda_{x,y}$, undulation wavelengths $(\lambda_x,\lambda_y)=(20,30)$ $\mu$m, the phase $\phi_x=\pi$, and the scaling factor of undulation amplitudes, $A_u=0$ \% (no undulation) to $100$ \% (strong undulation). The values of undulation amplitudes and wavelengths are chosen based on histology in mouse brain WM axons \citep{lee2019axial,lee20193dem} and its analysis in \ref{sec:app-undulation}.
Furthermore, the comparison of the 1-harmonic model (\ref{eq:D-undul-neuman-1h}) and the multi-harmonic model (\ref{eq:D-undul-neuman}) yields a prediction of fit parameters in \eqref{eq:D-undul-wide-1h}: 
\begin{equation} \label{eq:w-lambda}
w_0^2\lambda^2\simeq A_u^2(w_x^2\lambda_x^2+w_y^2\lambda_y^2)\,.
\end{equation}

To generate axons with randomly positioned beads, the radius $r(z)$ in \eqref{rho-gen}, varying along the $z$-axis, is calculated by convolving the random placement $n(z)$ of 1{\it d} barriers with a Gaussian kernel of width $\sigma$:
\begin{equation} \notag
    r(z) = r_0+r_1\cdot\left[ n(z) * \frac{e^{-z^2/(2\sigma^2)}}{\sqrt{2\pi\sigma^2}}\right]\,,
\end{equation}
where $r_0$ and $r_1$ are parameters determined by the mean cross-sectional area $A$ and radius's coefficient of variation $\text{CV}(r)$, defined as the ratio of the standard deviation (std) to the mean value:
\begin{equation} \label{eq:cv}
    A = \pi \cdot \langle r^2\rangle \,,\quad \text{CV}(r) = \frac{\text{std}(r)}{\langle r\rangle}\,.
\end{equation}

Based on histological observations in our previous studies \citep{lee2019axial,lee20193dem}, the bead width, determined by the FWHM of the Gaussian kernel, is fixed at $2\sqrt{2\ln2}\,\sigma=7$ $\mu$m, and the random bead placement $n(z)$ has a normally distributed inter bead distance $=5.70\pm2.88$ $\mu$m. The caliber variation $\text{CV}(r)$ varies  from 0 (no beads) to 0.4 (big beads) with a fixed mean cross-sectional area $A=\pi\cdot(0.5 \,\mu\text{m})^2$ for all axons.
}

\subsection{Realistic microstructure for simulations: \\ Intra-axonal space in WM}
Substrates were created by selecting 227 long myelinated axons (\figref{fig:IAS-shape}a-b), segmented from SEM images of a female mouse's genu of corpus callosum \citep{lee20193dem}.
The IAS mask was down-sampled into an isotropic resolution of (0.1 $\mu$m)$^3$ and aligned to the $z$-axis to control the orientation dispersion (\figref{fig:IAS-shape}c). The aligned axons were cropped at both ends to avoid oblique end faces, leading to axons of $\sim$18 $\mu$m in length \mpar{R4.1}\new{and of $\sim$1 $\mu$m in diameter}. More details were provided in our previous work \citep{lee20193dem}.

To quantify caliber variations, we estimated the equivalent circle radius $r$, defined as the radius of an equivalent circle with the same cross-sectional area perpendicular to axon's main axis \citep{west2016gratio,lee20193dem}. Further, we calculated the radius's coefficient of variation, $\text{CV}(r)$ \new{in \eqref{eq:cv}}.

To evaluate undulations of the axonal shape, we calculated $\langle \delta w^2\rangle$ based on the realistic axonal skeleton and \eqref{eq:w2} with $D_a = 2$ $\mu$m\textsuperscript{2}/ms for \new{$t = 1-200$} ms. Furthermore, we par\-a\-me\-tri\-zed the straightness of axons via the {\it sinuosity} (i.e., tortuosity factor in \citep{nilsson2012undulation})%
, defined in \ref{sec:app-undulation} as the ratio of the curvilinear length (along the axonal skeleton) to the straight-line distance between the two ends\new{, \eqref{eq:sinuosity}}\mpar{R2.15}. \new{Sinuosity is a representation of the axon undulation:}\mpar{R3.3} The larger the sinuosity, the larger the undulation\new{, cf. \eqref{eq:sinuosity} for exact relation and \eqref{eq:sinuosiy-1-harmonic} for an approximate relation}. The axonal skeleton and its main axis were constructed as in \secref{sec:syn-fiber}, and the skeleton was smoothed by a Gaussian filter of width $\sigma=1$ $\mu$m along the axon's main direction. 

\new{Theoretically, it is possible to estimate the undulation wavelength $\lambda$ based on the Fourier transform of the axonal skeleton; however, in practice, it is difficult to do so due to the limited axonal length $\sim18$ $\mu$m. Therefore, we can only estimate individual axon's $\lambda$ by fitting the simulated wide pulse RD to the simplified 1-harmonic model in \eqref{eq:D-undul-wide-1h}.}

\subsection{MC simulations}
MC simulations of random walkers were implemented in CUDA C++ in a continuous space within the 3{\it d} micro-geometry. When a random walker encounters a boundary, the original step gets canceled and another direction is chosen randomly until the resulting step would not cross any boundaries (equal-step-length random leap) \citep{xing2013randomleap,fieremans2018phantom}. The implementation of the equal-step-length random leap effectively repels walkers away from the \new{membrane} by a distance on the order of the step size, leading to small biases in the axon radius (\ref{sec:app-radius}). 
The correction of the axon radius in \eqref{eq:dr-3d} was thus applied to Eqs.~(\ref{eq:reff}), (\ref{eq:K}) and (\ref{r62}) to calculate theoretical predictions of $\langle r^2\rangle_v$, \new{$\langle r^4\rangle_v$} and $K_\infty$. The top and bottom faces of each axon are extended by reflected copies to avoid geometrical discontinuity in simulations.

\new{\subsubsection{Narrow pulse sequence}}
\new{For the narrow pulse pulsed-gradient sequence, simulations in synthetic undulating axons with randomly positioned beads are performed using $2\times$10$^5$} random walkers per axon diffusing over \new{$1\times$10$^6$} steps with a duration $\delta t = 2\times$10$^{-4}$ ms and a length $\sqrt{6D_0\delta t}=0.049$ $\mu$m, where the intrinsic diffusivity $D_0=2$ $\mu$m$^2$/ms is taken to agree with recent experiments \citep{novikov2018rotinv,veraart2019highb,dhital2019ias}; maximal diffusion time is \new{200} ms. For simulations in realistic IAS, \new{$2.27\times$10$^7$} random walkers in total diffuse over \new{$1\times$10$^6$} steps with a duration $2\times$10$^{-4}$ ms and a length $0.049$ $\mu$m; maximal diffusion time is \new{200} ms. Total calculation time is $\sim2$ days on a single NVIDIA Tesla V100 GPU at the NYU Langone Health BigPurple high-performance-computing cluster.

RD and RK of each axon were estimated based on cumulants of the diffusion displacement perpendicular to the axon by using \eqref{eq:D-K-def}.

The effective radius $r_\text{eff}$ estimated based on simulation results was obtained by fitting simulated $D_\perp(t)$ to \eqref{eq:Dt} for \new{$t=160-200$} ms. Similarly, the RK in $t\to\infty$ limit, $K_\infty$, estimated based on simulation results was \new{obtained by fitting simulated $K_\perp(t)$ to \eqref{eq:Kt} for $t = 160-200$ ms}. The above estimations based on simulations were further compared with theoretical predictions in Eqs.~(\ref{eq:reff}) and (\ref{eq:K}).

\new{
\subsubsection{Wide pulse sequence}
For the wide pulse pulsed-gradient sequence, simulations in synthetic undulating axons with randomly positioned beads are performed using $2\times10^5$ random walkers per axon diffusing over $1\times10^6$ steps with a duration $2\times10^{-4}$ ms and a length $0.049$ $\mu$m. For simulations in realistic IAS, $2.27\times10^7$ random walkers in total diffuse over $1\times10^6$ steps with a duration $2\times10^{-4}$ and a length $0.049$ $\mu$m. For simulations in all geometries, the diffusion time $t=1-100$ ms is the same as the gradient pulse width $\delta$ ($=t$), and diffusion signals are calculated based on the accumulated diffusional phase for ten b-values $=0.2-2$ ms/$\mu$m$^2$ along x- and y-axes transverse to axon's main direction ($z$-axis). Total calculation time is $\sim$2 days.

RD and RK of each axon were estimated by fitting the cumulant expansion to simulated pulsed-gradient diffusion signals $S$ \citep{jensen2005dki}:
\begin{equation}\label{eq:cum}
\ln S = -bD_\perp + \frac{1}{6}(bD_\perp)^2K_\perp + {\cal O}(b^3)\,,
\end{equation}
with diffusion weighting $b$ in \eqref{eq:b-value} for the wide pulse.

To evaluate axon undulations, we fitted the 1-harmonic model in \eqref{eq:D-undul-wide-1h} to simulated RD, $D_\perp(t,\delta)$ at $t=\delta=10-100$ ms, and calculated the effective radius due to undulations, $r_\text{und}$ in \eqref{eq:reff-wide-undul}, using fit parameters ($w_0$, $\lambda$).

The effective radius for wide pulse, $r_\text{eff, WP}$, was estimated based on \eqref{eq:reff-wide} and simulated RD, $D_\perp(t,\delta)$. The estimated $r_\text{eff, WP}$ was further compared with theoretical predictions, $r_\text{cal}$ in \eqref{r62} for caliber regime and $r_\text{und}$ in \eqref{eq:reff-wide-undul} for undulation regime.
}

\new{
\subsubsection{Directionally averaged signal}
To observe the directionally averaged signal decay versus diffusion weighting $b$, simulations in realistic IAS are performed applying $2.27\times10^7$ random walkers in total diffusing over $1\times10^6$ steps with a duration $2\times10^{-4}$ and a length $0.049$ $\mu$m. The diffusion and gradient pulse width $(t,\delta)=(20,7.1)$, $(30,13)$ and $(50.9,35.1)$ ms are chosen to match the experiments performed on animal 16.4T MR scanner (Bruker BioSpin), clinical 3T MR scanner (Siemens Prisma), and Siemens Connectome 3T MR scanner \citep{veraart2020highb}. Diffusion signals are calculated based on the accumulated diffusional phase for each of 18 b-values $=16-100$ ms/$\mu$m$^2$ along 30 uniformly distributed directions for each b-shell. Total calculation time is $\sim2$ days.

Directionally averaged signal $\overline{S}_i$ for individual axon was calculated by averaging diffusion signals of all directions for each b-shell, and the volume-weighted sum of all axons was also calculated:
\begin{equation} \label{eq:S-all-axons}
    \overline{S} = \frac{\sum\limits_i f_i\cdot \overline{S}_i}{\sum\limits_i f_i}\,.
\end{equation}

The RD, $D_\perp(t,\delta)$, was estimated by fitting \eqref{eq:S-sm} to simulated $\overline{S}_i$ and $\overline{S}$ over the range $b=55-100$ ms/$\mu$m$^2$ for individual axons and for all axons, and the effective radius of wide pulse, $r_\text{eff, WP}$, was calculated based on estimated $D_\perp$ and \eqref{eq:reff-wide} and compared with theoretical predictions in \eqref{r62}.
}

\subsection{Data and code availability}
The SEM data and IAS segmentation can be downloaded on the cai2r website (http://cai2r.net/resources/software), and the simulation codes can be downloaded on our github page (https://github.com/NYU-DiffusionMRI).

\begin{figure*}[htb!]\centering
\includegraphics[width=0.99\textwidth]{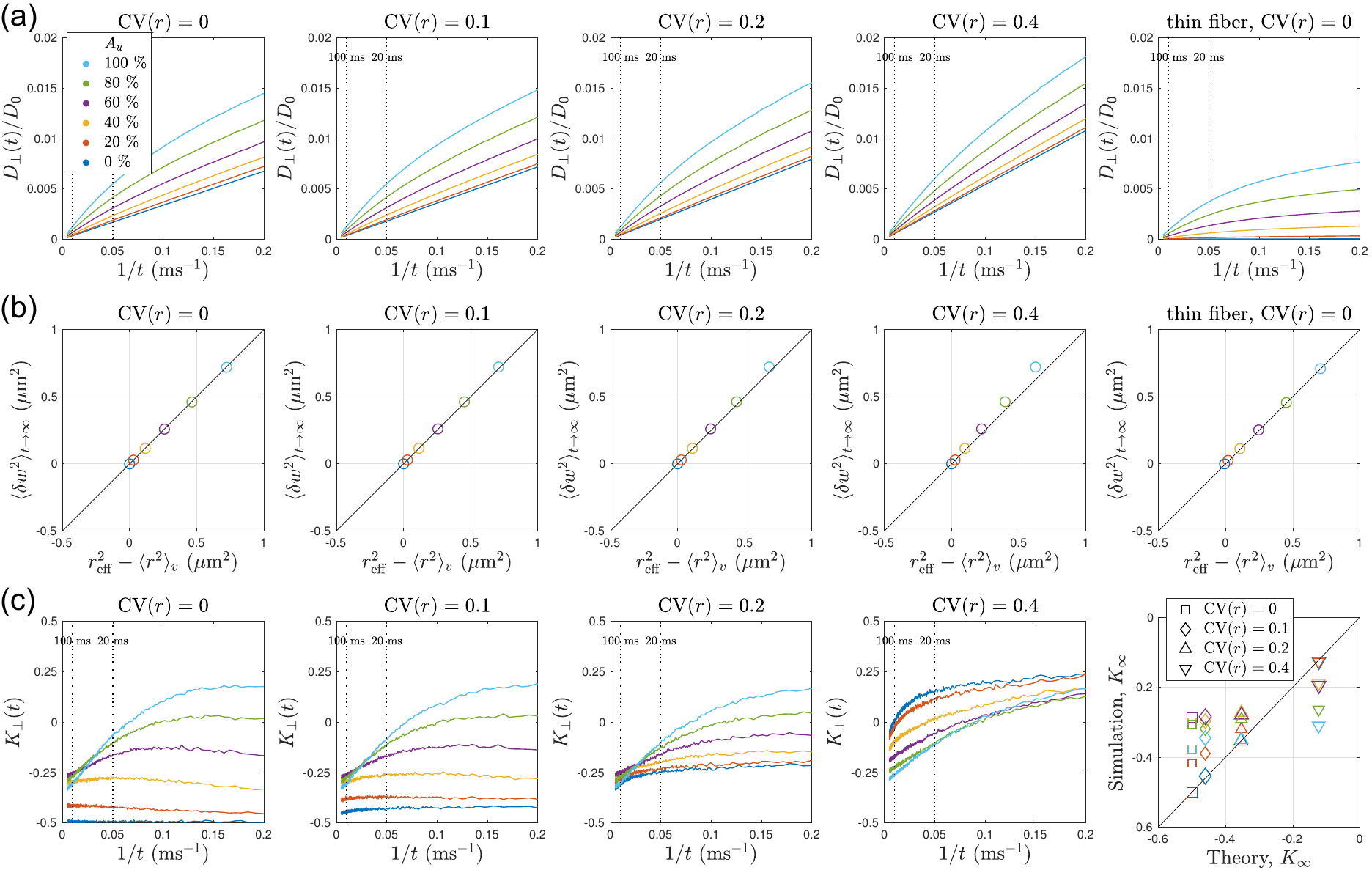}
\caption[]{\mnote{R1.1\\R2.13}\new{Undulations and caliber variations provide similar contributions to $D_\perp(t)$, $K_\perp(t)$ and $r_\text{eff}$ in the narrow pulse limit: 
synthetic axons with randomly positioned beads and sinusoidal undulations (\figref{fig:bead-shape}). 
(a) Simulated RD, $D_\perp(t)$, scales as $1/t$ at long times. (b) Effective radius, $r_\text{eff}^2$ fitted based on \eqref{eq:Dt} and simulations, deviates from the theoretical prediction $\langle r^2\rangle_v$ in \eqref{eq:reff} with the consideration of caliber variation only. This deviation can be explained by the contribution of axon undulation $\langle \delta w^2\rangle_{t\to\infty}$ at long times, estimated based on the axon skeleton and Eqs.~(\ref{eq:w2}) and (\ref{eq:dreff}). (c) For the axon with no caliber variations and undulations ($\text{CV}(r)=0$ and $A_u=0$\%, dark blue line in the left panel), simulated RK,  $K_\perp(t)$, is constant over time ($\sim-1/2$). For other axons, however, $K_\perp(t)$ scales as $1/t$ at long times. Furthermore, RK in $t\to\infty$ limit, $K_\infty$ fitted based on \eqref{eq:Kt} and simulations, is consistent with theoretical prediction of \eqref{eq:K} when the axon undulation is negligible ($A_u=0$\%, dark blue data points in the right panel).}}
\label{fig:bead-result}
\end{figure*}

\begin{figure*}[htb!]\centering
\includegraphics[width=0.99\textwidth]{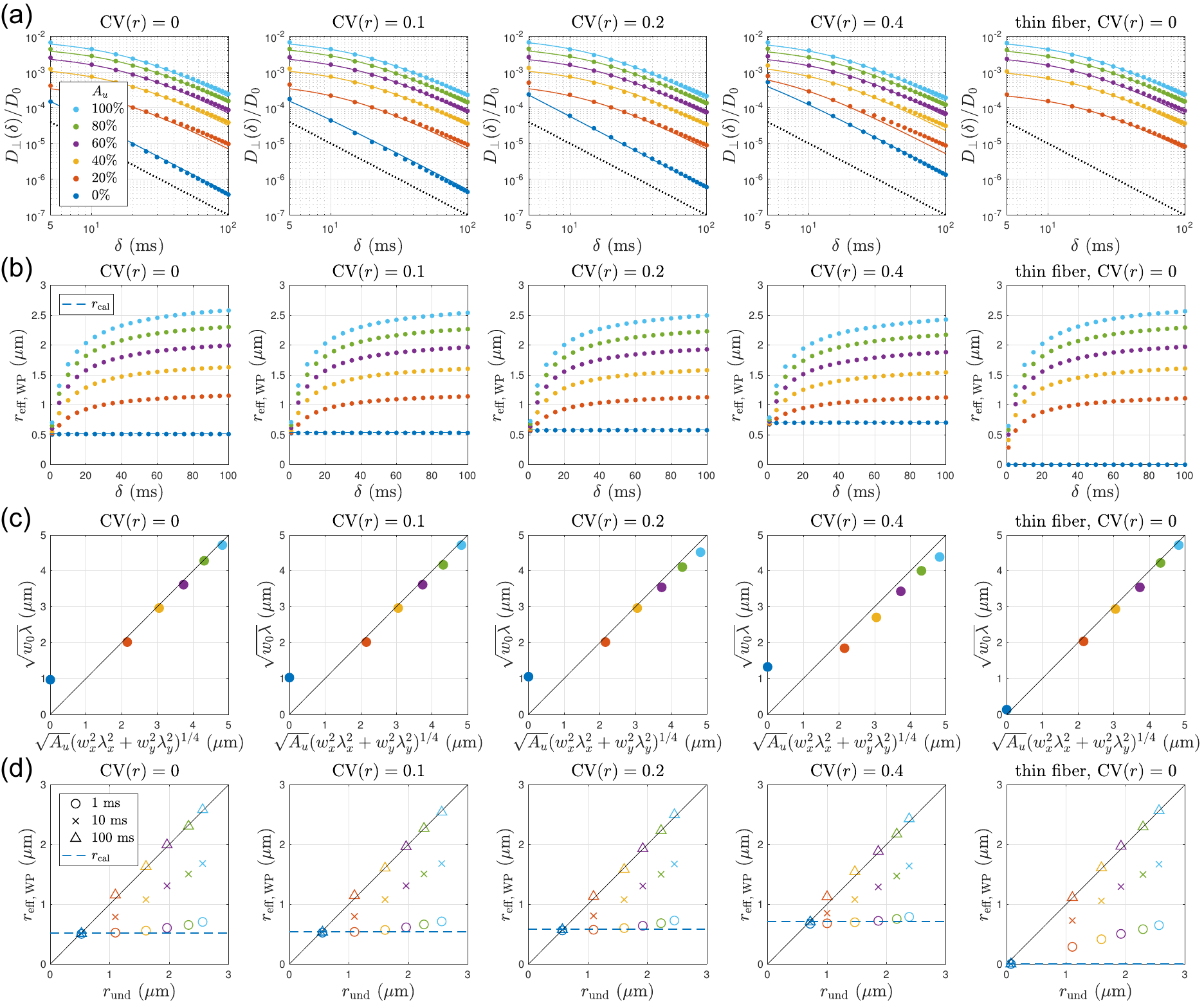}
\caption[]{\mnote{R1.1\\R1.3\\R3.1\\R4.2}\new{
Undulation contribution to $D_\perp(\delta)$ and $r_\text{eff, WP}$ dominates that of axon caliber variation for wide pulses ($t=\delta$): synthetic axons with randomly positioned beads and sinusoidal undulations. 
(a) Simulated RD, $D_\perp(\delta)$, in axons with caliber variations and undulations tuned via $\text{CV}(r)=0-0.4$ and $A_u=0-100$ \% respectively. To evaluate the effect of undulations on RD, we also simulate in undulating thin axons with no caliber variations. The solid lines are fits to the 1-harmonic undulation model in \eqref{eq:D-undul-wide-1h}. The black dotted line is a reference line $\propto \delta^{-2}$.  (b) For undulating axons ($A_u>0$\%), the effective radius of wide pulse sequences, $r_\text{eff, WP}$ estimated based on \eqref{eq:reff-wide} and simulations, is much larger than the theoretical prediction $r_\text{cal}$ in \eqref{r62} (dark blue dashed line) at long times. (c) The fit parameter $\sqrt{w_0\lambda}$ of the 1-harmonic undulation model in \eqref{eq:D-undul-wide-1h} is consistent with its theoretical value $\sqrt{A_u}(w_x^2\lambda_x^2+w_y^2\lambda_y^2)^{1/4}$ in \eqref{eq:w-lambda}, except for axons with no undulations ($A_u=0$\%).  (d) Comparison of $r_\text{eff, WP}$ at $\delta=1,10,100$ ms with theoretical predictions in {\it caliber regime}, $r_\text{cal}$ in \eqref{r62} (dark blue dashed line), and in {\it undulation regime}, $r_\text{und}$ in \eqref{eq:reff-wide-undul} (solid black line).}}
\label{fig:lissajous-wide}
\end{figure*}

\begin{figure*}[ht!]\centering
\includegraphics[width=0.99\textwidth]{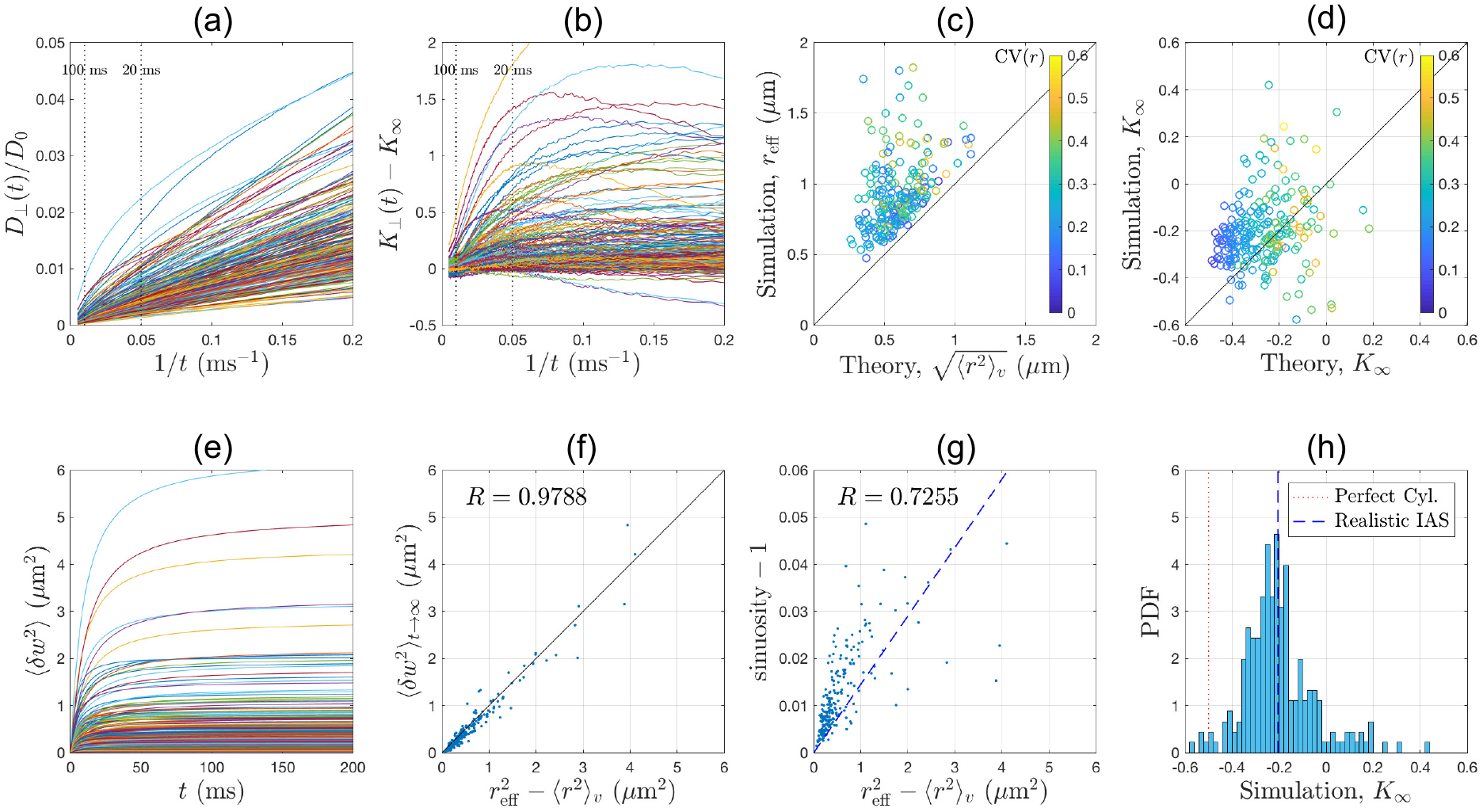}
\caption[]{Realistic intra-axonal space, \new{narrow pulse limit: undulations and caliber variations contribute similarly, as in \figref{fig:bead-result}}. 
(a-b) \new{Simulated RD and RK, $D_\perp(t)$ and $K_\perp(t)$, both scale as $1/t$ at long times for most axons. To observe the kurtosis time-dependence, we fitted \eqref{eq:Kt} to simulated $K(t)$ and subtracted the fitted $K_\infty$ (RK in $t\to\infty$ limit) from the simulation.} (c) Effective radius, $r_\text{eff}$ fitted based on \eqref{eq:Dt} from the simulations, is larger than theoretical predictions in \eqref{eq:reff} due to axon undulations, cf. \eqref{eq:dreff}. 
(d) $K_\infty$ fitted based on \eqref{eq:Kt} and simulations, is centered at about $-0.2$, notably larger (smaller in magnitude) than the $K_\infty$ = $-1/2$ for a perfectly straight cylinder. Markers in (c-d) are colored based on the value of the coefficient of variation of the radius, \eqref{eq:cv}.
(e) The second-order cumulant due to the axonal undulation, $\langle \delta w^2\rangle$ in \eqref{eq:w2}, increases with diffusion times $t$, and gradually approaches a constant at long times. (f) The cumulant $\langle \delta w^2\rangle_{t\to\infty}$ at long times (approximated by the value at \new{$t=200$} ms) has a non-trivial contribution to the RD, leading to the discrepancy of the effective radius estimation, $r_\text{eff}^2-\langle r^2\rangle_v$, as predicted by \eqref{eq:dreff}.
(g) The bias of the effective radius estimate in (c), $r_\text{eff}^2-\langle r^2\rangle_v$, highly correlates with axon's sinuosity (\new{the linear fit, blue dashed line, yields $\lambda$ from \eqref{eq:dreff-sinuosity}}). (h) The histogram of $K_\infty$ based on fitting \eqref{eq:Kt} to the simulation results, i.e., the values of the vertical axis in (d).  $K_\infty$ for the realistic axonal shapes has an average $K_\infty\simeq -0.2$ (blue dashed line), as compared to $-1/2$ for a perfectly straight cylinder (red dotted line).}
\label{fig:IAS-result}
\end{figure*}

\section{Results}

\subsection{Diffusion transverse to \new{synthetic undulating axons with randomly positioned beads}}
For simulations \new{of narrow pulse sequence in synthetic undulating axons with randomly positioned} beads, the simulated time-dependent RD, $D_\perp(t)$ in \figref{fig:bead-result}a, scales as $1/t$ in accordance with \eqref{eq:Dt-narrow} for all axons. 
\new{Similarly, the simulated RK, $K_\perp(t)$ in \figref{fig:bead-result}c, also scales as $1/t$, in accord with \eqref{eq:Kt}.}

The effective radius of narrow pulse, $r_\text{eff}$ fitted based on \eqref{eq:Dt} and simulations in \figref{fig:bead-result}a, \new{deviates from the theoretical predictions given by \eqref{eq:reff} with the consideration of caliber variation only (\figref{fig:bead-result}b). This deviation can be explained by the contribution of axon undulations $\langle 
\delta w^2\rangle_{t\to\infty}$ at long times, as predicted in Eqs.~(\ref{eq:w2}) and (\ref{eq:dreff}). Similarly, RK in $t\to\infty$ limit, $K_\infty$ fitted based on \eqref{eq:Kt} and simulations in \figref{fig:bead-result}c, agrees with the theoretical predictions given by \eqref{eq:K} when undulations are negligible ($A_u=0$ \%, the right panel in \figref{fig:bead-result}c).} 

\new{For simulations of wide pulse sequence in synthetic axons, the RD, $D_\perp(\delta)\equiv D_\perp(t,\delta)|_{t=\delta}$ in \figref{fig:lissajous-wide}a, scales as $1/\delta^2$ for axons with no undulations ($A_u=0$ \%) when $t=\delta$, consistent with the theoretical prediction in \eqref{eq:D-neuman} due to caliber variations. However, for axons with non-trivial undulations, $D_\perp(\delta)$ is much larger than the prediction in the caliber regime. Instead, the functional form of $D_\perp(\delta)$ can be explained by the simple 1-harmonic undulation model in \eqref{eq:D-undul-wide-1h}, indicating the dominant impact of undulations on the axon size estimation. The effect of undulations is further demonstrated by translating the simulated $D_\perp$ into the effective radius for wide pulses, $r_\text{eff, WP}$ in \eqref{eq:reff-wide}, which is much larger than the prediction of the caliber regime, $r_\text{cal}$ in \eqref{r62}, and increases dramatically with time $\delta$ (\figref{fig:lissajous-wide}b). 

Furthermore, we compare the fit parameters $\sqrt{w_0\lambda}$ based on $D_\perp$ in \figref{fig:lissajous-wide}a and 1-harmonic undulation model (\ref{eq:D-undul-wide-1h}) with the theoretical prediction in \eqref{eq:w-lambda} (\figref{fig:lissajous-wide}c); the fit parameters coincide with the theory except for axons with no undulations ($A_u=0$\%). Finally, we compare the estimate of $r_\text{eff, WP}$ at $\delta=1,10,100$ ms with predictions of caliber regime, $r_\text{cal}$, and of undulation regime, $r_\text{und}$ in \eqref{eq:reff-wide-undul} (\figref{fig:lissajous-wide}d). At very short time $\sim1$ ms, estimated $r_\text{eff, WP}$ coincides with $r_\text{cal}$ and is hardly influenced by undulations. However, at long time $\sim100$ ms, $r_\text{eff, WP}$ is consistent with $r_\text{und}$ and completely confounded by undulations. At relatively short time $\sim10$ ms, the axon size estimate falls into an intermediate regime and is partly biased by undulations. }

\subsection{Diffusion transverse to realistic axons}
For simulations \new{of narrow pulse sequences} in realistic IAS, simulated RD, $D_\perp(t)$ in \figref{fig:IAS-result}a, scales asymptotically as $1/t$ at long times \new{($t\gtrsim$ 20 ms)}\mpar{R2.17} in most axons, in \new{accordance} with \eqref{eq:Dt}. Similarly, simulated RK, $K_\perp(t)$ in \figref{fig:IAS-result}b, \new{also scales as $1/t$} at long times, in \mpar{R1.6}\new{accordance} with \eqref{eq:Kt}.

The effective radius \new{for narrow pulse,} $r_\text{eff}$ fitted based on \eqref{eq:Dt} and simulations in \figref{fig:IAS-result}a, is larger than the theoretical prediction of caliber variation $(\langle r^2\rangle_v)^{1/2}$ in \eqref{eq:reff}, as shown in \figref{fig:IAS-result}c. 
As we will discuss below, this discrepancy between the two estimates is due to the axonal undulation, confirming the undulations as an important mechanism for the apparent axonal diameter \citep{nilsson2012undulation,brabec2019undulation}.

The undulation $\langle \delta w^2\rangle$ based on \eqref{eq:w2} in a realistic axonal shape increases with diffusion time, and gradually approaches a constant at long time (\figref{fig:IAS-result}e). The bias in the effective radius estimation, $r_\text{eff}^2-\langle r^2\rangle_v$, can be solely explained  by the axonal undulation $\langle \delta w^2 \rangle_{t\to\infty}$ \new{at long times} (\figref{fig:IAS-result}f, \eqref{eq:dreff}) and highly correlates with the sinuosity (\figref{fig:IAS-result}g).

The RK in $t\to\infty$ limit, \new{$K_\infty$ fitted based on \eqref{eq:Kt} and} simulation results in \figref{fig:IAS-result}b, is centered around \new{$-0.20 \pm 0.15$} (Figures~\ref{fig:IAS-result}d and \ref{fig:IAS-result}h), different from the value $-1/2$ of a perfect cylinder \citep{burcaw2015meso}.

\begin{figure}[tb!]\centering
\includegraphics[width=0.475\textwidth]{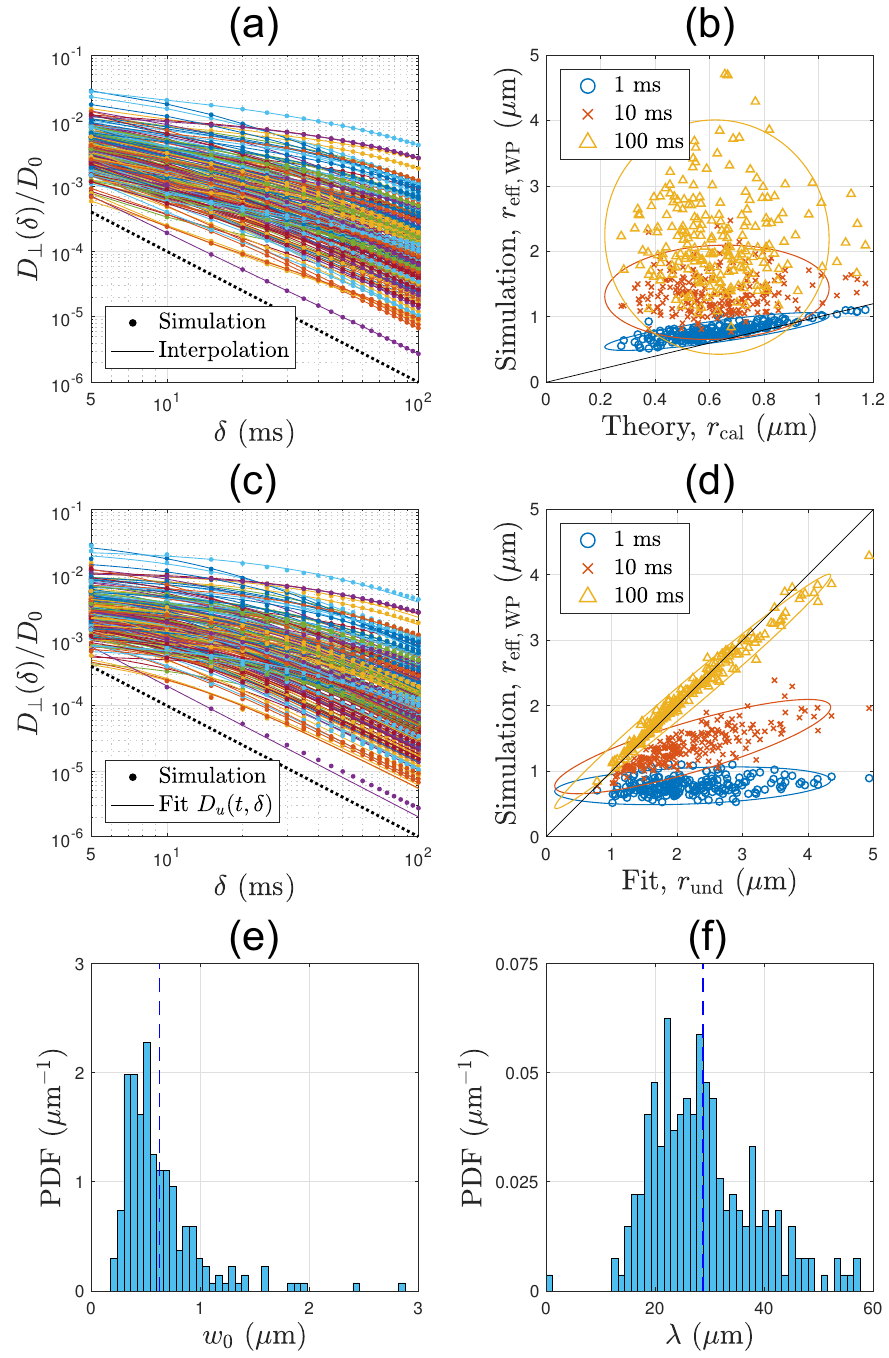}
\caption[]{\reversemarginpar\mnote{R1.3\\R3.1\\R4.2}\new{Undulations in realistic axons provide a dominant contribution over the caliber variations to $r_\text{eff, WP}$ in the wide-pulse limit, similar to \figref{fig:lissajous-wide}. 
Simulated results of realistic intra-axonal space for the wide pulse pulsed-gradient sequence ($t=\delta$). (a) Simulated RD, $D_\perp(\delta)$, with respect to $1/\delta$. The data points are simulation results, and solid lines are interpolations solely for visualization. The black dotted line is a reference line $\propto \delta^{-2}$.
(b) The effective radius of wide pulse, $r_\text{eff, WP}$ estimated based on \eqref{eq:reff-wide} and simulations, is larger than theoretical predictions, $r_\text{cal}$ in \eqref{r62}  practicall for all axons (for $r < 0.9$ $\mu$m), even at very short time $\delta\sim1$ ms. (c) Simulated $D_\perp(\delta)$ (data points) fitted to the 1-harmonic undulation model in \eqref{eq:D-undul-wide-1h} (solid lines). The black dotted line is a reference line $\propto \delta^{-2}$. (d) At long time $\delta\sim100$ ms, the estimated $r_\text{eff, WP}$ is consistent with the effective radius due to undulations, $r_\text{und}$ calculated based on \eqref{eq:reff-wide-undul} and fit parameters ($w_0$, $\lambda$) in \eqref{eq:D-undul-wide-1h}. And at relatively short time $\delta\sim10$ ms, the estimated $r_\text{eff, WP}$ is still partly biased due to undulations. This result indicates that the axon size estimation is confounded by undulations, unless at very short time $\delta\lesssim1$ ms. (e-f) Histograms of fit parameters of 1-harmonic undulation model: undulation amplitude $w_0$ and wavelength $\lambda$. The blue dashed lines indicate their mean values. }}
\label{fig:IAS-wide}
\end{figure}

\new{
For simulations of wide pulse sequences in realistic IAS, the RD, $D_\perp(\delta)$ in \figref{fig:IAS-wide}a, does not scale as $1/\delta^2$ at long times, incompatible with the prediction of caliber regime in \eqref{eq:D-neuman} when $t=\delta$. The effective radius of wide pulse, $r_\text{eff, WP}$ estimated based on \eqref{eq:reff-wide} and simulations in \figref{fig:IAS-wide}a, is larger than the theoretical prediction of caliber variations, $r_\text{cal}$ in \eqref{r62}, practically for all but the very largest axons with radii $\lesssim0.9$ $\mu$m at even very short time $\delta\sim1$ ms (\figref{fig:IAS-wide}b).

Furthermore, to evaluate the effect of undulation on RD, we fitted the simplified 1-harmonic undulation model in \eqref{eq:D-undul-wide-1h} to simulated $D_\perp(\delta)$ (\figref{fig:IAS-wide}c) and used the fit parameters to estimate the effective radius due to undulations, $r_\text{und}$ in \eqref{eq:reff-wide-undul}. At long time $\delta\sim100$ ms, the estimated $r_\text{eff, WP}$ coincides with the $r_\text{und}$, indicating that the axon size estimation is highly biased by undulations (\figref{fig:IAS-wide}d). Similarly, at relatively short time $\delta\sim10$ ms, $r_\text{eff, WP}$ is still partly biased due to undulations. This result demonstrates that the axon size estimation is inevitably confounded by undulations, unless performed at very short time $\delta\lesssim1$ ms. The fit parameters of 1-harmonic model suggest a rough scale of undulation amplitude $w_0\simeq 0.62\pm0.36$ $\mu$m and wavelength $\lambda\simeq29\pm9$ $\mu$m (\figref{fig:IAS-wide}e-f). 
}

\begin{figure*}[t!!]\centering
\includegraphics[width=0.75\textwidth]{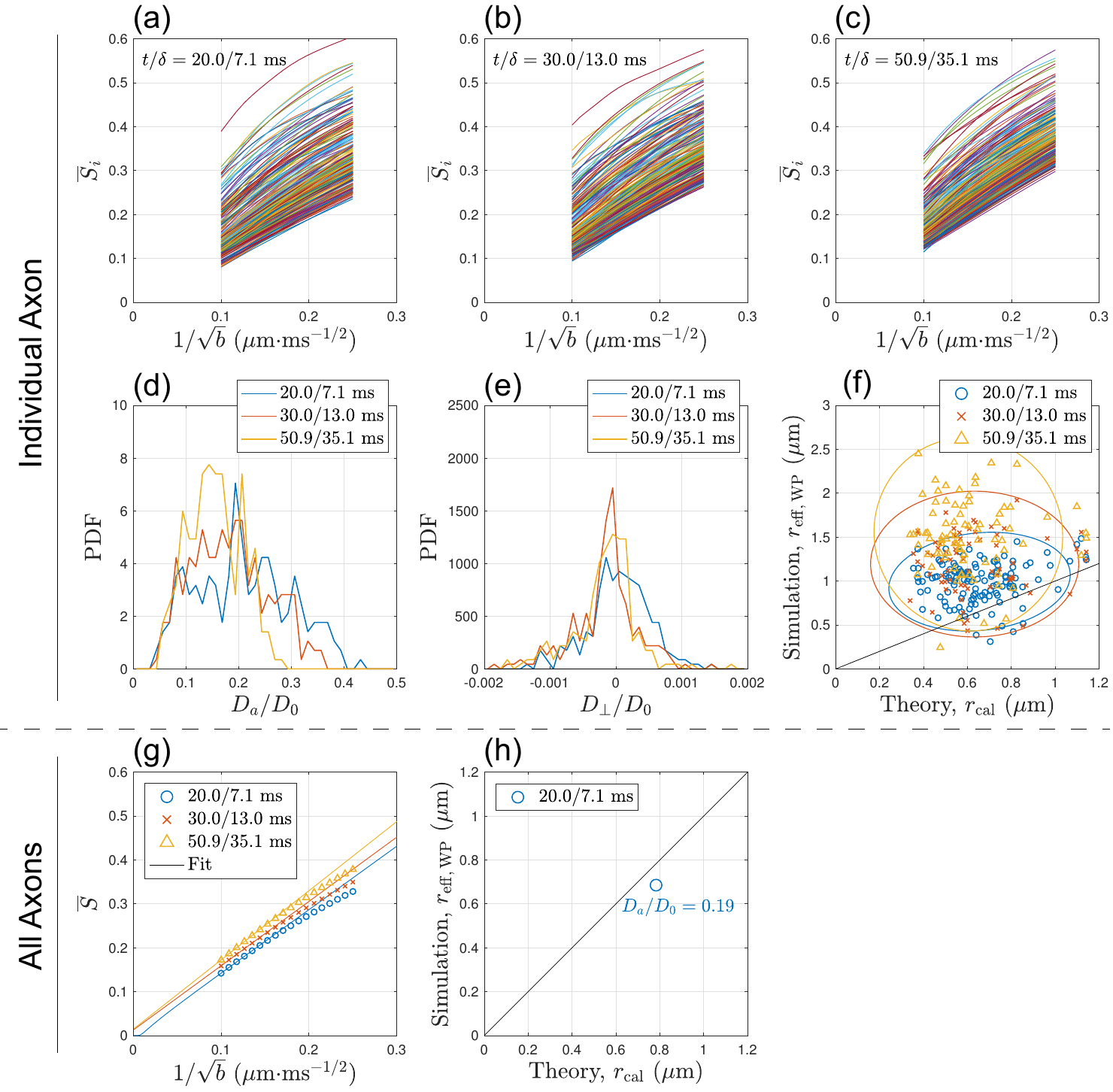}
\caption[]{\mnote{R1.3\\R3.1\\R4.2}\new{Factoring out the effect of undulations via directional averaging is most efficient at shortest times. 
Simulation results for realistic intra-axonal space and wide pulse sequence ($t=\delta$) with strong diffusion weighting $b=16-100$ ms/$\mu$m$^2$. (a-c) The directionally averaged diffusion signal $\overline{S}_i$ of individual axon scales roughly as $1/\sqrt{b}$ at high b-values for most axons. The fit of the full functional form in \eqref{eq:S-sm} to simulated signals yields the estimate of diffusivities along and transverse to axons, $D_a$ and $D_\perp$ respectively.  (d) The fitted diffusivity along axons, $D_a$, is smaller than the intrinsic diffusivity $D_0$. (e) The fitted diffusivity transverse to axons, $D_\perp$, is small with unphysical negative values in about half of axons. (f) Individual axon's effective radius of wide pulse, $r_\text{eff, WP}$ estimated based on \eqref{eq:reff-wide} and $D_\perp$ in (e), is larger than the theoretical prediction $r_\text{cal}$ in \eqref{r62}. (g) The directionally averaged diffusion signal $\overline{S}$ of all axons (volume-weighted sum) also scales as $1/\sqrt{b}$ at high b-values. For $t/\delta=30/13$ ms and $50.9/35.1$ ms, the $\overline{S}$ has positive signal intercepts as $1/\sqrt{b}\to0$, indicating unphysical negative diffusivities $D_\perp$ transverse to axons. (h) Effective radius of wide pulse of all axons, $r_\text{eff, WP}$, is consistent with the prediction $r_\text{cal}$ at short time $t/\delta=20/7.1$ ms, whereas the corresponding diffusivity $D_a$ along axons is unexpectedly low. At longer times $t/\delta=30/13$ ms and $50.9/35.1$ ms, the unphysical negative $D_\perp$ (indicated by positive signal intercept as $1/\sqrt{b}\to0$ in (g)) cannot provides an axon size estimation.}}
\label{fig:IAS-highb}
\end{figure*}

\new{
\subsection{Directionally averaged signal}
For simulations of wide pulse sequences of strong diffusion weightings in realistic IAS, the directionally averaged signal of individual axon, $\overline{S}_i$ in \figref{fig:IAS-highb}a-c, scales roughly as $1/\sqrt{b}$ in \eqref{eq:S-sm} at high b-values for most axons. 
We estimate the diffusivities along and transverse to axons, $D_a$ and $D_\perp$, based on \eqref{eq:S-sm} and simulation results in \figref{fig:IAS-highb}a-c. Individual axon's $D_a\sim0.2 \times D_0$ in \figref{fig:IAS-highb}d is unexpectedly low, and individual axon's $D_\perp$ in \figref{fig:IAS-highb}e is small with unphysical negative diffusivity values for about half of axons. Furthermore, individual axon's effective radius in the wide pulse limit, estimated based on \eqref{eq:reff-wide} and positive $D_\perp$ in \figref{fig:IAS-highb}e, is larger than the theoretical prediction $r_\text{cal}$ in \eqref{r62}.

Similarly, the directionally averaged signal combined from all axons, i.e., the volume-weighted sum $\overline{S}$ in \eqref{eq:S-all-axons}, also scales as $1/\sqrt{b}$ at high b-values (\figref{fig:IAS-highb}g). However, for $t/\delta=30/13$ ms and $50.9/35.1$ ms, the $\overline{S}$ has positive signal intercept as $1/\sqrt{b}\to0$, indicating unphysical negative diffusivities $D_\perp$ transverse to axons \citep{veraart2019highb,veraart2020highb}. Therefore, we can only estimate the effective radius of wide pulse, $r_\text{eff, WP}$, at relative short time $t/\delta=20/7.1$ ms, at which the estimated $r_\text{eff, WP}$ is consistent with the theoretical prediction $r_\text{cal}$. However, the corresponding fitted $D_a\simeq0.19\times D_0$ along axons is still unexpectedly low.
}

\section{Discussion}

In this work, we studied theoretically and numerically the interplay between axon caliber, caliber variations, and undulations, in synthetic and in realistic axons. 
The coarse-graining theory (\secref{sec:coh} and \ref{sec:app-averaging}) was developed for the narrow-pulse limit, when the diffusion gradient pulse width $\delta\ll t_D$.
\new{In this case, the caliber-variation contribution ($\sim r^2$) to the diffusivity transverse to axons is comparable to the undulation one ($\sim w_0^2$), i.e., $r\sim w_0\sim0.6$ $\mu$m (\figref{fig:IAS-wide}e). 

In the opposite, wide-pulse limit, the physics of the IAS signal attenuation is qualitatively different. The caliber-variation contribution ($\sim r^4$) to radial diffusivity becomes much smaller than that of the undulations ($\sim w_0^2\lambda^2$), because $r\ll\lambda\approx30$ $\mu$m (\figref{fig:IAS-wide}f).}
\new{Therefore, simply varying the gradient pulse width (while keeping $b$ constant) is not a feasible experimental probe to distinguish between the relative contributions of different confounding effects for the ADM, except at very short times $\sim1$ ms.} 

\new{Another approach of ADM is to partially factor out the undulation effect by taking the directionally averaged signals for each $b$-shell at relatively short times $\delta\lesssim10$ ms compared to $t_u\approx 11$ ms (defined after \eqref{eq:D-undul-wide-1h} for the typical values $\lambda=30$ $\mu$m, $D_a=2$ $\mu$m$^2$/ms). }

Below we  discuss the narrow-pulse limit (\secref{sec:narrow}), 
qualitative \new{and quantitative} differences arising in the wide-pulse limit (\secref{sec:wide}), \new{the directionally averaged signal (\secref{sec:discuss-sm})}, and 
interplay with the diffusion in the extra-axonal space (\secref{sec:discuss-inter-time}).

\subsection{Narrow pulse limit}
\label{sec:narrow}

Based on the cumulant expansion, the dMRI signal can be expressed as \eqref{eq:cum},
where $b\simeq (g\delta)^2\cdot t$.
We first discuss the $D_\perp(t)$ term in the narrow pulse limit, and then consider when the $K_\perp$ term becomes important. 

\subsubsection{Diffusion coefficient and \texorpdfstring{$r_\text{eff}$}{TEXT}}

Simulation results \new{of narrow pulse sequence in synthetic undulating axons with randomly positioned} beads show that axons of the same mean cross-sectional area can have very different dMRI-measured effective radius of narrow pulse, $r_\text{eff}$ in \eqref{eq:Dt}, depending on the strength of beads, $\text{CV}(r)$, \new{and strength of undulations, $A_u$. In \figref{fig:bead-result}b, dMRI-measured $r_\text{eff}$ overestimates the radius of undulating axons even with the consideration of caliber variations in \eqref{eq:reff}; this discrepancy, $r_\text{eff}^2-\langle r^2\rangle_v$, can be explained by the contribution of undulations $\langle \delta w^2\rangle_{t\to\infty}$ at long times in Eqs.~(\ref{eq:w2}) and (\ref{eq:dreff}). }


Similarly, in realistic IAS (\figref{fig:IAS-result}c), the $r_\text{eff}$ estimate, \new{fitted based on \eqref{eq:Dt} and simulation results in \figref{fig:IAS-result}a,} is larger than the estimate based on theory that includes only the caliber variations in \eqref{eq:reff}. 
The remaining discrepancy, $r_\text{eff}^2-\langle r^2\rangle_v$, is caused by the axonal undulation $\langle \delta w^2 \rangle_{t\to\infty}$ (\eqref{eq:dreff}, \figref{fig:IAS-result}f), and correlates with the axonal straightness/sinuosity (\eqref{eq:dreff-sinuosity}, \ref{sec:app-undulation-sinuosity}). \figref{fig:IAS-result}g shows that the larger the sinuosity (i.e., the larger the undulation), the bigger the discrepancy. This observation also coincides with the prediction of the 
toy model in \citep{brabec2019undulation}. 

The bias in effective radius estimation due to the undulation depends on the chosen diffusion time range. 
The longest diffusion time used to fit \eqref{eq:Dt} potentially increases the discrepancy $r_\text{eff}^2-\langle r^2\rangle_v$ since $\langle \delta w^2\rangle$ grows with diffusion times in \figref{fig:IAS-result}e. In other words, at long times (diffusion length $L_d(t)$ $\sim$ undulation wavelength $\lambda$ defined in \ref{sec:app-undulation-sinuosity}), the radius estimation will be biased due to the undulation the most.  

In particular, at sufficiently long times (e.g., $t\sim100$ ms), diffusion length along the axon, $L_d(t)=\sqrt{2D_at} \simeq 20$ $\mu$m, is comparable with the undulation wavelength \new{$\lambda\simeq 26$ $\mu$m}, estimated based on the slope in \figref{fig:IAS-result}g and the 1-harmonic model in \eqref{eq:dreff-sinuosity}, \ref{sec:app-undulation-sinuosity}.
Therefore, for such long $t$, the radius estimation can be highly biased due to the undulation. In contrast, at intermediate times (e.g., $t\sim5$ ms), the diffusion length along the axon, $L_d(t)\simeq 4.5$ $\mu$m, is much shorter than the undulation wavelength, and at the same time the diffusion length (without the restrictions) transverse to axons, 
$L_d^\perp(t)=\sqrt{4D_0t} \approx 6.3\,\mathrm{\mu m} \gg r\sim 1\, \mathrm{\mu m}$,  is still long enough  for \eqref{eq:Dt} to be applicable; in this case, the radius estimation based on \eqref{eq:reff} is less biased.

We find that, at long times ($t\gtrsim$100 ms), $\sim$50\% of $r_\text{eff}^2$ is contributed by the caliber variation, i.e., $\langle r^2\rangle_v$ calculated based on \eqref{eq:reff}, and the remaining $\sim$50\% is contributed by the undulation, i.e., $\langle \delta w^2\rangle_{t\to\infty}$ in \eqref{eq:w2} and \eqref{eq:dreff}.

Unfortunately, our EM segmentation was limited to fairly short axonal segments. We believe that quantifying the amplitude and wavelength for axonal undulations based on larger field-of-view $3d$ EM reconstructions will be  important for assessing the relevance of undulations  on the measured IAS characteristics. 

\cite{sepehrband2016adm} estimated the inner axonal diameter of the ex vivo mouse brain by using PGSE with diffusion time $t_\text{max}$ = 28 ms and ultra-high diffusion gradient strength $G_\text{max}$ = 1350 mT/m, and still overestimated the diameter index, $2\sqrt{\langle r^2\rangle_v}$ in \eqref{eq:reff} for narrow pulse and $2r_\text{cal}$ in \eqref{r62} for wide pulse, based on the histology by a factor of $\sim 2$. This bias can be partly explained by not including the undulation effect in the model.

\subsubsection{When is radial kurtosis important?}

To evaluate the importance of the kurtosis term, consider the cumulant expansion (\ref{eq:cum}). The $\sim b^2$ term is of the same order as the $\sim b$ one when a strong diffusion gradient is applied:
\begin{equation*}
g\gtrsim g^*=\sqrt{\frac{24}{|K_\infty|}} \cdot \frac{1}{r\cdot \delta }\,.
\end{equation*}
For example, if $\delta=$ 10 ms, $r=3$ $\mu$m (thick axons), and $K_\infty=-0.2$, $g^*= 0.3651$ ($\mu$m$\cdot$ms)$^{-1}$, corresponding to $G= 1365$ mT/m. When the applied diffusion gradient is of this order of magnitude, the kurtosis contribution becomes crucial;  furthermore, the effect of axonal shape and diameter distribution on the value $K_\infty$ (different from $-1/2$) becomes important. 

Based on simulations in realistic IAS, the intra-axonal signal is shown to have a non-negligible kurtosis $K_\infty\approx-0.2$ transverse to axons (Figures~\ref{fig:IAS-result}d and \ref{fig:IAS-result}h). Its value is different from that of a prefect cylinder, $K_\infty = -1/2$ \citep{burcaw2015meso}, and needs to be considered in axonal diameter measurements using dMRI, e.g., for large axons in spinal cord \citep{assaf2008axcaliber,duval2015spinal,benjamini2016wmadm}, with strong diffusion gradients applied \citep{duval2015spinal,sepehrband2016adm}, and for IAS metabolites \citep{ronen2014cc,palombo2016metabolite}. 

\subsection{Wide pulse limit}
\label{sec:wide}

\new{

\subsubsection{Contributions of caliber and undulations to the radial diffusivity}
\label{sec:disc-RD}

In synthetic undulating axons with randomly positioned beads, the RD time-dependence $D_\perp(t,\delta)$ when $t=\delta$  does not scale as $1/\delta^2$ predicted by \eqref{eq:D-neuman} due to caliber variations (\figref{fig:lissajous-wide}a). Rather, it is explained by the simple 1-harmonic model for $\delta\gtrsim t_u\sim11$ ms, corresponding to the undulation wavelengths $\lambda_x/\lambda_y=20/30$ $\mu$m. The bias of the 1-harmonic model at shorter $\delta\lesssim t_u$ is caused by the RD contribution from caliber variations and by fitting this simplified model to $D_\perp(\delta)$ contributed by undulations of 2-harmonics in \eqref{eq:lissajous}. Furthermore, the effective radius of wide pulse, $r_\text{eff, WP}$ in \eqref{eq:reff-wide}, exceeds the prediction $r_\text{cal}$ in \eqref{r62} for undulating axons except at very short times $\delta\lesssim1$ ms (\figref{fig:lissajous-wide}b). Instead, the effective radius due to undulations, $r_\text{und}$ in \eqref{eq:reff-wide-undul}, coincides with the estimated $r_\text{eff, WP}$ at long times $\delta\sim100$ ms (\figref{fig:lissajous-wide}d), indicating that the axon ``size'' estimation is inevitably dominated by axon undulations at clinical times $t$ and $\delta$. 

The competition of caliber and undulation regimes is also shown via fit parameters of the undulation model (\ref{eq:D-undul-wide-1h}) (\figref{fig:lissajous-wide}c). The fitted $\sqrt{w_0\lambda}$ is consistent with the theory (\ref{eq:w-lambda}) for axons with non-trivial undulations. However, in non-undulating axons, the fitted value is much larger than the theoretical prediction ($w_0\lambda=0$) since the axon caliber is misinterpreted as the undulation length scale due to similar functional forms in Eqs.~(\ref{eq:D-neuman}) and (\ref{eq:D-undul-neuman-1h}).
}

\new{
Similarly, our simulations in realistic axon shape (\figref{fig:IAS-wide}) demonstrate that, for pulsed-gradient sequence of finite pulse width, the functional form of RD time-dependence cannot be properly described by the contribution of caliber variations in \eqref{eq:D-neuman} at times $\delta\gtrsim10$ ms; instead, it is well explained by the simplified 1-harmonic undulation model in \eqref{eq:D-undul-wide-1h}. Indeed, the undulation of realistic axons is composed of multiple harmonics (\figref{fig:appendix-undulation-simulation}c-d), whose contributions to RD have exactly the same $1/\delta^2$-functional form only at times $\delta\gg t_u\approx11$ ms for the undulation wavelength $\lambda\approx29$ $\mu$m (\eqref{eq:D-undul-neuman} and \figref{fig:IAS-wide}f); in other words, when $\delta\gg t_u$, RD contributions from all harmonics degenerate into the same $1/\delta^2$-functional form, and the overall RD contribution from multiple harmonics has exactly the same functional form as of 1-harmonic case.
Therefore, the 1-harmonic model deviates from simulation results at shorter times $\delta\lesssim t_u$ (\figref{fig:IAS-wide}c), whereas the fitting results for $\delta\gg t_u$ still provides reasonable estimates of undulation metrics: The estimated amplitude $\sim0.62$ $\mu$m in \figref{fig:IAS-wide}e is consistent with the analysis of axonal shapes in \figref{fig:appendix-undulation-simulation}c, and the estimated wavelength $\sim29$ $\mu$m in \figref{fig:IAS-wide}f almost coincides with the value $\sim26$ $\mu$m given by the correlation of sinuosity and the axon size discrepancy $r_\text{eff}^2-\langle r^2\rangle_v$ in \figref{fig:IAS-result}g.
}



\new{For the radius estimation in either narrow or wide pulse limit, axonal undulation has an essential contribution to RD. For example, in the narrow pulse limit} at typical diffusion time $t = 50$ ms, the RD due to undulations is $\langle \delta w^2\rangle/4t \simeq 2.3\times10^{-3}$ $\mu$m$^2$/ms averaged over all axons 
in our EM sample (\figref{fig:appendix-undulation-simulation}a), comparable to the RD due to the caliber variation: $\langle r^2\rangle_v/4t = 2.5\times10^{-3}$ $\mu$m$^2$/ms \new{(\figref{fig:IAS-result}a)}.
\new{
Further, in the wide pulse limit, the RD due to caliber variations can be estimated based on \eqref{eq:D-neuman} 
and $\langle r^4\rangle_v\sim (0.78$ $\mu$m$)^4$, and the RD due to undulations can be estimated based on \eqref{eq:D-undul-wide-1h} and $w_0/\lambda\sim 0.62/29$ $\mu$m (\figref{fig:IAS-result}e-f). At typical $t = 50$ ms and diffusion pulse width $\delta = 15$ ms, the RD due to caliber variations is 
$D_\text{cal, WP} \simeq 2.5\times10^{-5}$ $\mu$m$^2$/ms, 
much smaller than the RD contributed by undulations, $D_u\simeq1.4\times10^{-3}$ $\mu$m$^2$/ms.
}

Finally, for diffusion-weighted MR spectroscopy, the intra-axonal markers (e.g., {\it N}-acetyl aspartate + {\it N}-acetyl aspartyl glutamate, tNAA) have a smaller diffusivity (i.e., longer $t_D$), and the measurement could potentially fall into the narrow pulse limit. 
However, the measurements performed so far 
still fall into the wide pulse limit. For example, the diffusivity of tNAA, $D_\text{tNAA} = 0.51$ $\mu$m$^2$/ms in the \new{anterior body of} human corpus callosum, 
where axons have radii  $r< 3$ $\mu$m, yields the correlation time $t_D=r^2/D_\text{tNAA}<18$ ms, much shorter than the applied gradient pulse width $\delta = 42$ ms in \citep{ronen2014cc}.

\new{
In addition to diffusion transverse to axons, the undulation effect on diffusion along axons (\ref{sec:app-undulation-axial}) has been in vivo observed in healthy subjects: Using the planar water mobility filter with diffusion weighting $\leq 5.4$ ms/$\mu$m$^2$, \citet{dhital2019ias} measured the axial diffusivity and trace of diffusion tensor in single bundle voxels, respectively approaching the intra-axonal diffusivity $D_\parallel$ in the principal fiber bundle direction and the intra-axonal diffusivity $D_a$ factoring out fiber dispersion and axonal undulations. Based on the in vivo measurements ($D_\parallel/D_a\sim2.25/2.39$ $\mu$m$^2$/ms) and theory in \eqref{eq:D-ax-und}, we have $w_0/\lambda\approx 0.055$, corresponding to $\lambda\approx 11$ $\mu$m if $w_0\sim0.62$ $\mu$m. This estimation is smaller than ours $\lambda\approx29$ $\mu$m in the mouse brain.
}

\new{
\subsubsection{When is radial kurtosis important?}
\label{sec:disc-RK}

Let us now estimate the contribution of higher order $g^4$ term; we focus on an axon with constant cross-section for simplicity. As it follows from \eqref{eq:S-wide-pulse} and \citep{lee2018rd}, the $g^4$ term becomes comparable with the $g^2$ term when the following condition is satisfied:
\begin{equation*}
g\gtrsim g^* = \sqrt{\frac{c_1}{c_2}}\cdot \frac{D_0}{r^3}\,,
\end{equation*}
where $\sqrt{c_1/c_2}\simeq 8.14$ for a perfectly straight cylinder, cf. \eqref{c1c2}. 
For example, if $r=3\,\mu$m and $D_0=2\,\mu$m\textsuperscript{2}/ms, \new{$g^*=0.6030$} ($\mu$m$\cdot$ ms)$^{-1}$, corresponding to \new{$G=2254$} mT/m (for thinner axons the critical gradient will be even larger); when the applied diffusion gradient is of this order of magnitude, the higher-order terms become comparable to the $g^2$ term, and will strongly bias the ADM results (in particular, the negative $g^4$ term will lead to radius overestimation). 

For the directionally averaged signal in \eqref{eq:S-sm}, the higher-order term ${\cal O}(b^2)$ (related to kurtosis) in the exponential term is typically ignored, cf., e.g., \cite{veraart2020highb}. 
This approximation can be now justified for ADM in brain based on the above analysis, as the critical gradient $g^*$ is much higher than the gradient strength available on clinical scanners, and even on most animal scanners.}

\new{\subsection{Directionally averaged signal}
\label{sec:discuss-sm}
}

\new{
To partly factor out the undulation effect on diffusion transverse to axons, simulated dMRI signals at strong diffusion weighting $b$ are directionally averaged (\figref{fig:IAS-highb}a-c). For simulations of individual axons, the signal $\overline{S}_i$ scales as $1/\sqrt{b}$ for most axons, and the $\overline{S}_i$ with negative signal intercepts at $1/\sqrt{b}\to0$ has biologically plausible estimates of RD $D_\perp>0$ \citep{veraart2020highb}, and the corresponding effective radius $r_\text{eff, WP}$ of wide pulse in \eqref{eq:reff-wide} (\figref{fig:IAS-highb}e-f): At short times ($t/\delta=20/7.1$ ms from \citep{veraart2020highb} on animal scanner), estimated $r_\text{eff, WP}$ is slightly larger with the prediction $(\langle r^2\rangle_v)^{1/4}$ in \eqref{r62}, whereas at longer times ($t/\delta=30/13$ ms, $50.9/35.1$ ms on human scanners \citep{veraart2019highb,veraart2020highb}), $r_\text{eff, WP}$ remarkably overestimates the axon size. Although the estimated radii at short times have reasonable values, the estimated diffusivity $D_a$ along axons is unexpectedly low ($\sim 0.2\times D_0$), much smaller than the values reported in previous in vivo studies, $D_a\lesssim D_0$ \citep{novikov2018rotinv,veraart2019highb,dhital2019ias}. This low $D_a$ value is potentially caused by ignoring the local fluctuations of $D_a(l)$ varying along individual axons, resulted from caliber variations along axons \citep{jacobs1935fjeq,novikov2014meso,lee2019axial}. 
Indeed, the saddle point estimate (\ref{eq:S-sm}) yielding $\overline{S}\sim 1/\sqrt{D_a}$ overemphasizes the parts of the axon with the smaller local $D_a(l)$, since  the average over the axon $\langle 1/\sqrt{D_a(l)}\rangle_l > 1/\sqrt{\langle D_a(l)\rangle_l}$. 
}

\new{
Similarly, the signal $\overline{S}$ of all axons (volume-weighted sum) scales as $1/\sqrt{b}$ (\figref{fig:IAS-highb}g-h). In particular, at short times, the $\overline{S}$ has a negative signal intercept and a reasonable estimate of $r_\text{eff, WP}$, and again the estimated $D_a$ ($\sim0.19\times D_0$) is lower than expected due to caliber variations along individual axons and across multiple axons; however, at long times, the $\overline{S}$ has positive intercepts and unphysical $D_\perp<0$.
}

\cite{veraart2020highb} measured intra-axonal RD $D_\text{wide\,pulse}$ $\sim1.2 \times10^{-2}$ $\mu$m$^2$/ms in vivo in the human brain WM, based on the deviation from the $1/\sqrt{b}$ scaling in directionally averaged dMRI signals up to $b\lesssim 25$ ms/$\mu$m$^2$, with $t = 30$ ms and $\delta = 13$ ms. Assuming that axonal undulation in the human brain WM is comparable to that observed in our EM sample, we can estimate the RD due to the undulation \new{$D_u \simeq 2.3\times10^{-3}$ $\mu$m$^2$/ms, accounting for at most $\sim20$ \% of measured intra-axonal RD if the undulation effect is not fully factored out in directionally averaged signals.}

\subsection{Contributions to the total radial kurtosis from intra- and extra-axonal spaces} 
\label{sec:discuss-inter-time}

To evaluate the contribution of intra-axonal kurtosis to the overall radial kurtosis with the presence of the extra-axonal space, we considered a fiber bundle consisting of multiple axons and an extra-axonal space as described in \secref{sec:theory-incoherent}. At long diffusion times, $D_i\propto 1/t$ is much smaller than $D_e$, i.e., $D_i\ll D_e$. Substituting into \eqref{eq:D-K-intra-extra}, we obtain 
\begin{subequations} \begin{align*}
\overline{D}(t)&\simeq (1-f)\cdot D_e(t)\,,\\
\overline{K}(t)&\simeq \frac{3f}{1-f} + \frac{K_e(t)}{1-f}\,. 
\end{align*} \end{subequations}
In particular, because it is the values $D_i^2 K_i$ that add up for the 4th-order cumulant, the IAS kurtosis contribution  to the overall kurtosis can be neglected at long $t$; the cross-term (diffusion variance) yields the first term in the right-hand side, and the extra-axonal kurtosis provides the main (time-dependent) correction, that decreases with time as $(\ln t)/t$ \citep{burcaw2015meso}. 
In other words, at long times, the overall kurtosis $\overline{K}$ solely depends on the intra-axonal volume fraction $f$ \citep{fieremans2010karger,fieremans2011dki,jensen2010dki} and the extra-axonal kurtosis $K_e(t)$.  

However, in the intermediate-time regime $t\sim r^2/D_0$,  the diffusivities are of the same order, $D_i \sim D_e$,  and therefore the intra-axonal kurtosis $K_i$ has a non-negligible contribution to the overall kurtosis $\overline{K}$. In this case, the intra-axonal kurtosis value becomes important for the overall radial kurtosis. Practically, this is relevant only at rather short diffusion times. 

\new{\subsection{Clinical significance} \mpar{R3.2}
The fact that dMRI is sensitive to caliber variations (e.g., beadings) and axonal undulations, as validated in both \citep{budde2010bead,brabec2019undulation} and this study, suggests potential applications of monitoring axon pathology. For example, the increase of axonal undulations has been observed at post-mortem acutely following traumatic brain injury (TBI) in humans \citep{tang2012microtubule}. Furthermore, axonal varicosity (beading) is a pathological change observed after TBI \citep{tang2012microtubule,johnson2013tbi} and ischemic injury to WM axons \citep{garthwaite1999ischemic}. Therefore, the so-called axonal ``diameter'' mapping, measured by using dMRI, could be sensitive to the increase of undulations in TBI patients as well as the increase of varicosities in TBI and ischemic stroke patients.}

\subsection{Limitations}
In this study, we only focused on the IAS of myelinated axons in WM. However, other structures, such as unmyelinated axons and the extra-axonal space, are also important and need to be considered, with finite water residence time in them taken into account. \new{\mpar{R3.1}Therefore, we choose not to discuss the signal-to-noised ratio required for ADM since extra-axonal signals are not simulated. In addition, the axonal diameter distribution is not considered in this study since dMRI measurements do not have sensitivity to probe signals contributed by axons in diameters $<2$ $\mu$m \citep{veraart2020highb}.}

Further, as discussed above, we only performed MC simulations for the pulsed-gradient sequence in narrow \new{and finite pulse widths}. The effect of coarse-graining on other commonly used sequences, such as the oscillating-gradient sequence, should be further considered to be able to compare with a range of existing measurements. We note, however, that the knowledge of full narrow-pulse $D(t)$ can be in principle used to predict the second-order cumulant for any sequence \cite[Sec.~2]{novikov2019note}, as we did here to derive the wide-pulse diffusivity for the undulations. Beyond the 2nd-order cumulant, full-scale MC simulations are needed for each gradient wave form. 

\section{Conclusions}
\new{
Numerical simulations, either in synthetic undulating axons with randomly positioned beads or in realistic axonal shapes from mouse brain EM, show that, in the narrow pulse limit, the inner diameter based on the RD is overestimated if the caliber variation and axonal undulation are both ignored, and the RK at long times is different from the value of perfectly straight cylinders. Furthermore, in the wide pulse limit, the contribution of undulations to RD time-dependence dominates and overshadows the contribution of caliber variations. In other words, conventional axonal ``diameter'' mapping is very sensitive to the strength of undulations, except at very short diffusion times. To release the requirement of very short diffusion times for an accurate ADM, we can factor out part of the undulation effect via a directional averaging of signals at strong diffusion weightings. To sum up, by decomposing salient features in the realistic axonal shape and recomposing their contributions to diffusion metrics, the theory and its numerical validation in this study builds a foundation for the biophysical interpretation and experimental implementation of axonal microstructure mapping.
}


\section{Acknowledgements}
We would like to thank the 
NYULH Bigpurple High-Perfor\-mance-Computing Center for numerical computations on the cluster, and Valerij Kiselev  for discussions. Research was supported by the National Institute of Neurological Disorders and Stroke of the NIH under award number R21 NS081230 and R01 NS088040, and by the National Institute of Biomedical Imaging and Bioengineering (NIBIB) of the NIH under award number U01 EB026996, and was performed at the Center of Advanced Imaging Innovation and Research (CAI2R, www.cai2r.net), a Biomedical Technology Resource Center supported by NIBIB under award number P41 EB017183. SJ acknowledges financial support for a sabbatical at NYU from Aar\-hus University Research Foundation (AUFF), the Lund\-beck Foun\-da\-tion (R291-2017-4375), and Augustinus Fonden (18-1456). SJ is also supported by the Danish National Research Foundation (CFIN), and the Danish Ministry of Science, Innovation, and Education (MINDLab).

\figref{fig:IAS-shape} is adapted with permission from \cite{lee20193dem}
Copyright 2019 Springer. 

\appendix
\setcounter{figure}{0}
\section{Coherent averaging along an axon with caliber variation}
\label{sec:app-averaging}
The diffusion signal measured by narrow-pulse monopolar PGSE is given by \new{(p.~340, eq.~[6.28], \citealt*{callaghan1991book})}\mpar{R2.18}
\begin{equation} \label{eq:S-dMRI}
S({\bf q}, t) = \frac{1}{V}\int \rho({\bf x})\, {\cal G}({\bf x}, {\bf x'}; t) e^{-i{\bf q}\cdot({\bf x'}-{\bf x})} d{\bf x}d{\bf x'}\,,
\end{equation}
where $\rho({\bf x})$ is the shape of the pore (confining or not and in any spatial dimensionality, with volume $V$), \eqref{eq:poreshape}, 
and ${\cal G}$ is the diffusion propagator, i.e., the probability of a spin at $\bf x$ diffusing to $\bf x'$ during time $t$. 

Consider the diffusion within a fully confining pore. In the long time limit ($t\to\infty$), the spin loses its memory of the initial position $\bf x$, and has an equal probability of being anywhere in the pore \new{(p.~378, eq.~[7.14], \citealt*{callaghan1991book})}\mpar{R2.18}:
\begin{equation} \label{eq:G-callaghan}
{\cal G}({\bf x}, {\bf x'}; t)\simeq \frac{1}{V}\rho({\bf x'})\,,\quad t\to\infty\,.
\end{equation}
Note that the propagator (\ref{eq:G-callaghan}) is  properly normalized (as the probability density for any $t$), 
\begin{equation} \label{eq:norm}
\int\! d{\bf x'} \, {\cal G}({\bf x}, {\bf x'}; t) = 1\,,
\end{equation} 
ensuring the normalization of $S({\bf q}, t)|_{q=0}= 1$ in \eqref{eq:S-dMRI}. 
Substituting \eqref{eq:G-callaghan} into \eqref{eq:S-dMRI}, the well-known diffusion diffraction result is obtained \new{(p.~378, eq.~[7.14], \citealt*{callaghan1991book})}\mpar{R2.18}:
\begin{equation} \label{eq:S-long-time}
S({\bf q}, t)|_{t\to\infty} \simeq \frac{|\rho({\bf q})|^2}{V^2}\,, \quad 
\rho({\bf q}) = \int\! \rho({\bf x})\,  e^{-i{\bf q}\cdot{\bf x}}d{\bf x} \,.
\end{equation}
For a 2{\it d} confined pore, \new{${\bf x}=(x_1,x_2)$}, the diffusion wave vector ${\bf q} = {\bf q}_\perp$ is parallel to the $x$-$y$ plane, and \eqref{eq:S-long-time} yields \eqref{eq:S}.

Consider now a segment of a 3{\it d} axon, of length $L_z$ smaller than the diffusion length $L_d(t)= \sqrt{2D_a t}$ along its direction. 
Suppose that we artificially confine all molecules to diffuse within this segment (by closing its ends), and measure diffusion  ${\bf q} = {\bf q}_\perp$ transverse to its axis. \eqref{eq:S-long-time} now applies for this 3{\it d} confining ``pore", and yields  
\begin{equation} \label{eq:segment}
S({\bf q}, t)|_{t\to\infty} \simeq \frac{| \langle\rho({\bf q}_\perp)\rangle_z |^2}{A^2}\,, \quad A = \frac{V}{L_z} \,, 
\end{equation}
where we expressed the result in terms of the  {\it two-dimensional} form factor 
\begin{equation}
\langle\rho({\bf q}_\perp)\rangle_z \equiv \frac1L_z \int\! d^3{\bf x} \, \rho({\bf x})\,  e^{-i{\bf q}_\perp \cdot{\bf x}}
= \int\! d^2{\bf x}_\perp \bar\rho({\bf x}_\perp) \, e^{-i{\bf q}_\perp \cdot{\bf x}_\perp} 
\end{equation}
of the average axonal cross-section 
\begin{equation} \label{eq:rhoav}
\bar\rho({\bf x}_\perp) \equiv \langle\rho({\bf x})\rangle_z = \frac{1}{L_z}\int\! dz\, \rho({\bf x}) \,.
\end{equation} 
For clarity, here we made explicit the dimensions of integration
It is important to note that the longitudinal average over the axonal cross-sections in \eqref{eq:segment} is {\it coherent}, i.e., we average the 2{\it d} form factors with their phases, and only then take the absolute value. 

To extend the above intuition onto a real (non-confining) axon, we note that the longer $L_d(t)$ we take, the longer confining axonal segment $L_z < L_d(t)$ we are allowed to consider in our gedanken setting above. If the structural disorder in the axonal caliber variations has a finite correlation length $l_c$ in the $z$ direction, then 
any segment with $L_z \gg l_c$ will ``look" like any other, in the sense that their values of $\langle\rho({\bf q}_\perp)\rangle_z$ will asymptotically become all equal to each other. This is the result of the self-averaging due to the coarse-graining by diffusion. 



Now one can realize that the notion of a closed segment in this argument has been auxiliary, and the role of the confinement window is played by the diffusion length $L_d(t)$, as it is the ``soft" envelope that is really confining the molecules. 
Hence, we propose the following ansatz for the diffusion propagator at long times:
\begin{equation} \label{eq:G-ansatz}
{\cal G}({\bf x},{\bf x'};t) = \frac{1}{\tilde{A}(z,t)}\rho({\bf x'}) G_0(z-z';t)\,,
\end{equation}
where $\tilde{A}(z,t)$ is a normalization factor (an average cross-section within the domain $\sim L_d(t)$ around the point ${\bf x}$) found from \eqref{eq:norm}, and $G_0$ is a 1{\it d} Gaussian diffusion propagator with a diffusivity $D^\parallel_\infty = D_a(t)|_{t\to\infty}$ along the axon. The propagator in \eqref{eq:G-ansatz} effectively factorizes the diffusion within the axon into Gaussian diffusion $G_0$ at long distances $\sim L_d(t)$ along the axon, and confined diffusion $\rho({\bf x'})$ at short distances $\sim r$ (typical radius) transverse to the axon. This approximate factorization only works when the separation of scales $L_d(t)\gg l_c$ and $L_d^\perp(t)\gg r$ has been reached, with the former ensuring Gaussian diffusion along the axon, and the later ensuring confined diffusion transverse to the axon.
It can be viewed as an extension of the ansatz of \cite{mitra1992propagator}, originally developed for the fully-connected 2{\it d} or 3{\it d} pore space, onto the quasi-1{\it d} situation where the motion is unconfined in one dimension and confined by hard walls in the others. 
Substituting \eqref{eq:G-ansatz} into \eqref{eq:S-dMRI}, approximating $\tilde{A}(z,t)$ with the mean cross-sectional area $A$, and employing ${\bf q}={\bf q_\perp}$, we obtain
\begin{equation}\label{eq:S-qperp-t}
S({\bf q}_\perp,t) \simeq \frac{1}{VA}\int \frac{dq_\parallel}{2\pi} e^{-D_\infty^\parallel q_\parallel^2 t} \left|\rho({\bf q}_\perp,q_\parallel)\right|^2\,. 
\end{equation} 
\eqref{eq:S-qperp-t} can be viewed as a rigorous definition of the coherent average in \eqref{eq:S-cg}, where the 3{\it d} form factor is being coarse-grained (Gaussian-filtered) 
$\rho({\bf q}_\perp,q_\parallel) \to \rho({\bf q}_\perp,q_\parallel) e^{-D_\infty^\parallel q_\parallel^2 t/2}$ over the window $\sim L_d(t)$, 
cf. also the discussion after Eq.~[10] of \citep{novikov2014meso}. Of course, in the $t\to \infty$ limit, the Gaussian filter selects $|q_\parallel| \lesssim 1/L_d(t) \to 0$, acting as the Dirac delta function
$\delta_D(q_\parallel)$, thus making \eqref{eq:S-qperp-t} equivalent to \eqref{eq:segment}. 

\begin{figure}[t!!!]\centering
\includegraphics[width=0.42\textwidth]{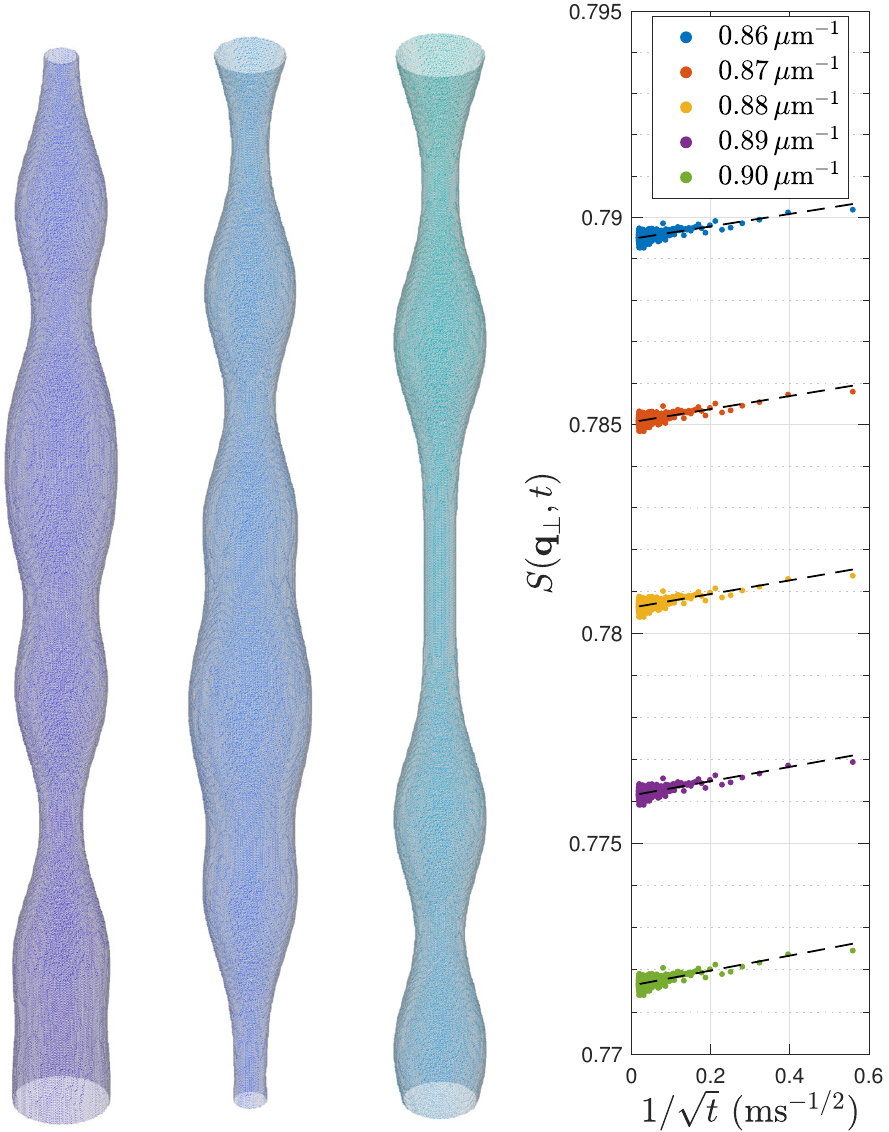}
\caption{Simulated signals $S({\bf q_\perp},t)$ in a cylinder of randomly distributed beads. Diffusion gradient is applied transverse to the axon, with $|{\bf q_\perp}|$ = 0.86-0.90 $\mu$m$^{-1}$. For $t$ = 3.2-3200 ms, $S({\bf q_\perp},t)$ scales as $1/\sqrt{t}$, consistent with the prediction in \eqref{eq:S-cg-correction}, validating the ansatz for the diffusion propagator in \eqref{eq:G-ansatz}.}
\label{fig:appendix-signal-correction}
\end{figure}

To validate our ansatz (\ref{eq:G-ansatz}) for the propagator and its consequence for the signal (\ref{eq:S-qperp-t}), we explore the finite-$t$ corrections to its $t\to\infty$ limit, \eqref{eq:segment}. 
For that, we study the effect of the deviations 
 $\delta\rho({\bf x}) = \rho({\bf x}) - \bar\rho({\bf x}_\perp)$ from the average shape, \eqref{eq:rhoav}. 
Substituting the corresponding 
\[
\rho({\bf q}) = \langle \rho({\bf q}_\perp)\rangle_z \cdot 2\pi \delta_D(q_\parallel) + \delta\rho({\bf q}_\perp, q_\parallel)
\]
into \eqref{eq:S-qperp-t} and using $\delta_D^2(q_\parallel) = \delta_D(q_\parallel) \cdot L_z/2\pi$, 
the cross-terms vanish since $\delta \rho|_{q_\parallel=0}\equiv 0$, and we obtain
\begin{multline} \label{eq:S-cg-correction}
S({\bf q}_\perp,t)\simeq \frac{1}{A^2}\left|\langle \rho({\bf q}_\perp)\rangle_z \right|^2 \\+ \frac{1}{VA}\int \frac{dq_\parallel}{2\pi} e^{-D_\infty^\parallel q_\parallel^2 t} \left|\delta\rho({\bf q}_\perp,q_\parallel)\right|^2\,,
\end{multline}
where the first right-hand-side term is the full along-axon coherent average, \eqref{eq:segment}, and the second term is a signal correction due to the microstructural inhomogeneity along the axon. 

For the $1d$ short-range disorder in the positions of axonal swellings, compatible with the along-axon diffusivity scaling as 
$\sim t^{-1/2}$ at long $t$ \citep{novikov2014meso,fieremans2016time},  the power spectrum of the density fluctuation $\left|\delta\rho({\bf q}_\perp,q_\parallel)\right|^2$ has a plateau at $q_\parallel\to 0$, leading to a signal correction term $\delta S({\bf q_\perp},t)$ $\propto 1/\sqrt{t}$ at long $t$ 
according to \eqref{eq:S-cg-correction}. Note that this scaling is not for the diffusivity, as in (\cite{novikov2014meso}), but for the {\it signal} itself. 
It provides the 1{\it d} generalization of the $[l_c/L_d(t)]^d \sim t^{-d/2}$ scaling \citep[Section 1.6]{novikov2019note} due to the correction to the free propagator following from the ansatz of \cite{mitra1992propagator} in dimensions $d>1$. 


In \figref{fig:appendix-signal-correction}, we demonstrate this $1/\sqrt{t}$ signal correction via numerical simulations in an artificially designed cylinder ($L_z$ = 200 $\mu$m, $A$ = $\pi\times$ (1 $\mu$m)\textsuperscript{2}) composed of randomly distributed beads, with distance between beads = 5.3 $\pm$ 2.9 $\mu$m and bead width = 4 $\mu$m. 1$\times$10\textsuperscript{7} random walkers diffuse within this beaded cylinder with the same parameter settings as in \new{the narrow pulse regime in} the main text, except $|{\bf q_\perp}|$ = 0.86-0.9 $\mu$m$^{-1}$ and $t$ = 3.2-3200 ms. The simulated signal has a very small $1/\sqrt{t}$-dependence ($<$ 0.2\% signal change in \figref{fig:appendix-signal-correction}), showing that, practically, one can simply use the full coherent averaging approximation, \eqref{eq:segment}, in our result, \eqref{eq:S-cg} of the main text.

\setcounter{figure}{0}
\new{\section{Diffusional kurtosis transverse to the perfectly straight cylinder}
\label{sec:app-cyl}

\begin{figure}[tb!]\centering
\includegraphics[width=0.45\textwidth]{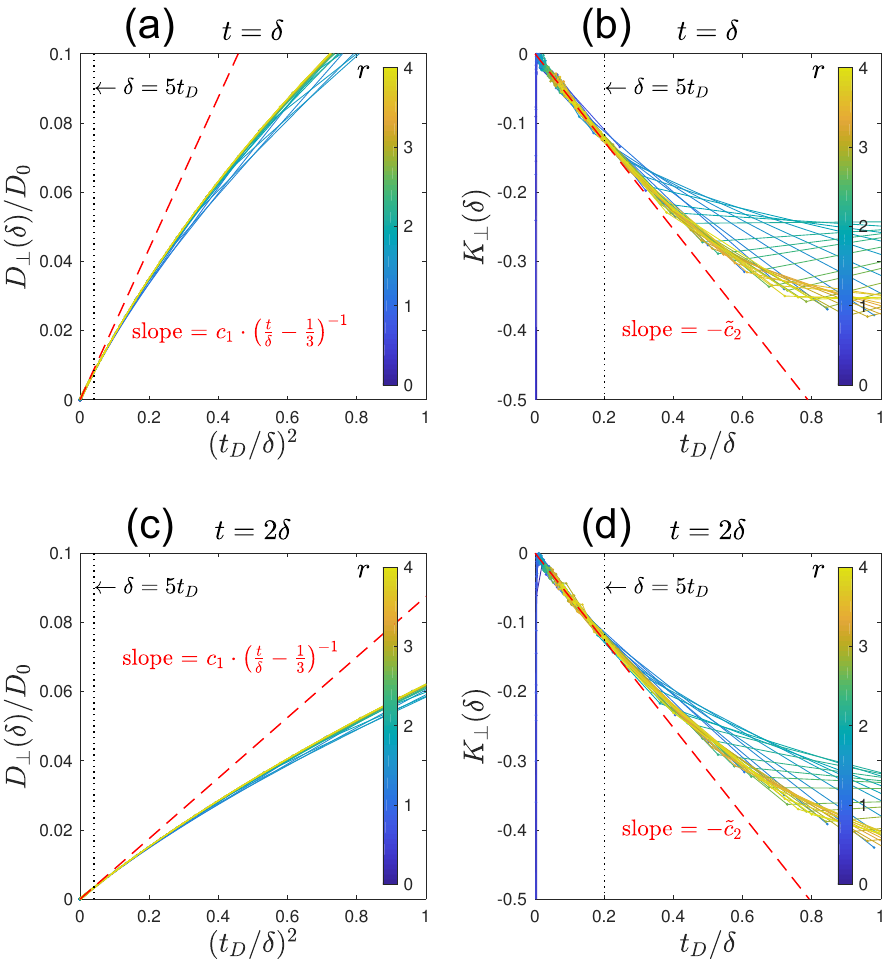}
\caption[]{\new{Simulation results for diffusion inside straight cylinders with circular cross-section of radii $r=0.1-4$ $\mu$m, for the wide pulse pulsed-gradient sequence with $t=\delta$ in (a-b) and $t=2\delta$ in (c-d). In the wide pulse limit $\delta\gg t_D(r)=r^2/D_0$, we observe the universal scaling relations for radial diffusivity (a,c) and radial kurtosis (b,d), such that the curves for different radii begin to coincide. 
(a,c): Radial diffusivity $D_\perp(\delta)$ scales as $t_D^2(r)/\delta^2$, and the slope (red dashed line) asymptotically verifies the exact value $c_1$ from \eqref{c1c2}.
(b,d): Radial kurtosis $K_\perp(\delta)$ scales as $-t_D(r)/\delta$, and the slope (red dashed line) determines $\tilde{c}_2$ in \eqref{eq:K-wide}, and thereby $c_2$ in \eqref{eq:S-wide-pulse} and \eqref{c1c2}, based on \eqref{eq:c2-value}.} }
\label{fig:appendix-cyl}
\end{figure}

In the narrow pulse limit ($\delta\ll t_D = r^2/D_0$), diffusion kurtosis transverse to a perfectly straight cylinder is a constant $K_\perp\simeq-$0.5 at long times \citep{burcaw2015meso}. On the other hand, in the wide pulse limit ($\delta\gg t_D$), the radial kurtosis has a non-trivial time-dependence given by \citep{lee2018rd}
\begin{equation} \label{eq:K-wide}
    K_\perp(\delta)\simeq -\tilde{c}_2\cdot\frac{t_D}{\delta} \,,
\end{equation}
where $\tilde{c}_2>0$ is a proportionality constant. The purpose of this section is to numerically estimate the value of $\tilde{c}_2$, as well as of $c_2$ in \eqref{eq:S-wide-pulse}.

To do so, we performed MC simulations of diffusion within perfectly straight cylinders of radii $r=0.1-4$ $\mu$m (aligned to the $z$-axis), applying $5\times10^5$ random walkers with $D_0=2$ $\mu$m$^2$/ms and $\delta t=2\times10^{-4}$ ms, and calculated diffusion signals of pulsed-gradient sequences with gradients applied transverse to cylinders (along $x$- and $y$-axes). To prevent the bias due to the equal-step-length random leap (\ref{sec:app-radius}) and pixelated/boxy geometry, {\it elastic collision} was implemented when a random walker encountering the curved surface of a cylinder. A gradient pulse width $\delta=1-100$ ms was applied, with diffusion time $t=\delta$ for the first simulation and $t=2\delta$ for the second one. Diffusion signals for five b-values = $0-1$ ms/$\mu$m$^2$ were simulated and fitted to the DKI representation in Eqs.~(\ref{eq:cum}) and (\ref{eq:b-value}) to calculate the radial diffusivity $D_\perp(t,\delta)$ and radial kurtosis $K_\perp(t,\delta)$.

As expected in  \eqref{eq:D-neuman}, simulated $D_\perp(\delta)$ scales linearly with $1/\delta^2$ for $\delta\gg t_D$ (\figref{fig:appendix-cyl}a,c). Similarly, as predicted by \eqref{eq:K-wide}, simulated $K_\perp(\delta)$ scales linearly with $1/\delta$ when $\delta\gg t_D$ (\figref{fig:appendix-cyl}b,d). Furthermore, the slope of this linear scaling provides an estimate of $\tilde{c}_2\approx0.63$. Substituting Eqs.~(\ref{eq:D-neuman}), (\ref{eq:b-value}) and (\ref{eq:K-wide}) into \eqref{eq:cum} and comparing with coefficients in \eqref{eq:S-wide-pulse}, we obtain
\begin{equation} \label{eq:c2-value}
    c_2 = \tfrac{1}{6}c_1^2\cdot \tilde{c}_2 \approx 0.0022\,.
\end{equation}
}

\setcounter{figure}{0}
\section{Axonal undulation and diffusion transverse to axons}
\label{sec:app-undulation}

\new{To quantify the effect of axonal undulations on diffusion metrics, we introduce the sinuosity, consider the case of a 1-harmonic undulation in the narrow pulse limit, and then generalize onto multiple undulation harmonics and derive the effects of non-narrow pulses.}

\begin{figure*}[tb!]\centering
\includegraphics[width=0.99\textwidth]{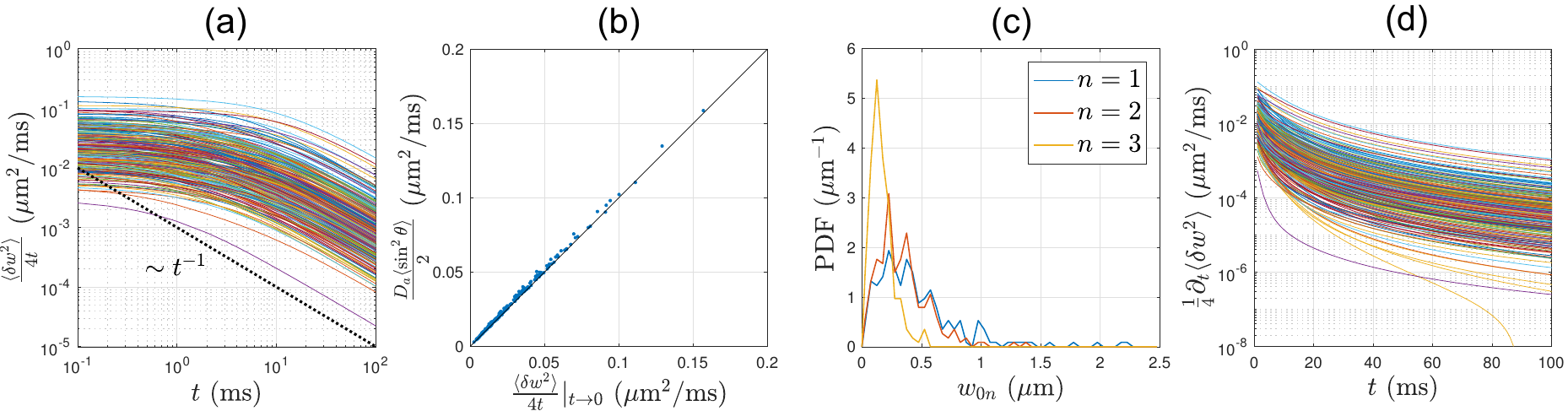}
\caption{(a) The time-dependence of the diffusivity $\langle \delta w^2\rangle/4t$ contributed by the undulation of the realistic IAS. The black dotted line is a reference line with a slope of $-$1, indicating a $1/t$-dependence at long times. (b) At short times, the diffusivity contributed by the undulation is the projection of the along-skeleton Gaussian diffusion onto the direction transverse to the main direction, as described in \eqref{eq:D-undulation-short-time}. (c) The histogram of the undulation amplitudes $w_{0n}$ of the first three harmonics ($n = 1$, $2$, $3$). (d) The time-dependence of the instantaneous diffusivity $\frac{1}{2d}\partial_t\langle \delta w^2\rangle$ in $d=2$ contributed by the undulation. In the semi-log plot, the instantaneous diffusivity  is not a straight line, i.e., it does not decrease mono-exponentially (\eqref{eq:Dinst-undulation}), indicating that the axonal undulation consists of more than one harmonic. }
\label{fig:appendix-undulation-simulation}
\end{figure*}

\subsection{Quantifying axonal undulations: Sinuosity}
\label{sec:app-undulation-sinuosity}
In \figref{fig:IAS-result}g, $r_\text{eff}^2-\langle r^2\rangle_v$ highly correlates with the sinuosity. Given that $r_\text{eff}^2-\langle r^2\rangle_v$ also correlates with the axonal undulation via $\eqref{eq:dreff}$ (\figref{fig:IAS-result}f), it is natural to consider the correlation between the undulation and the sinuosity.

Axonal sinuosity $\xi$ is defined as the ratio of the cuvilinear length $L=\int\! dl$ of the axonal skeleton to the Euclidean distance $L_z=\int\! dz$ of two ends:
\begin{subequations} \label{eq:sinuosity}\begin{align} \label{eq:sinuosity-exact}
\xi &\equiv \frac{L}{L_z} = \frac{1}{L_z} \int \sqrt{(dz)^2 + |d{\bf w}|^2}\\ \label{eq:sinuosity-small}
&\simeq 1+\frac{1}{2}\left\langle\left|\frac{d{\bf w}}{dz}\right|^2\right\rangle_z\,, \quad \left|\frac{d{\bf w}}{dz}\right|\ll 1\,,
\end{align} \end{subequations}
where ${\bf w}$ is defined in \figref{fig:appendix-undulation-shape}, and $\langle ...\rangle_z$ denotes the quantity averaged along the main axis. The deviation of axonal skeleton from the main axis in \figref{fig:appendix-undulation-shape}
correlates with the sinuosity via \eqref{eq:sinuosity}. In particular, \eqref{eq:sinuosity-small} is only applicable to axons with small undulations (i.e., $|d{\bf w}/dz|\ll 1$) and the sinuosity $\xi\gtrsim1$ (\figref{fig:IAS-result}g).

To better understand the meaning of \eqref{eq:sinuosity-small}, we consider a simple case of an axon with a sinusoidal undulation in one plane (the 1-harmonic model):
\begin{equation} \label{eq:1-harmonic}
{\bf w}=w_0\cos(kz) {\bf \hat{x}}\,,
\end{equation}
where $w_0$ is the undulation amplitude, $k=2\pi/\lambda$, and $\lambda$ is the undulation wavelength. 
\eqref{eq:sinuosity-small} yields 
\begin{equation} \label{eq:sinuosiy-1-harmonic}
\xi\simeq 1+\frac{1}{4}w_0^2 k^2 = 1+\left( {\pi w_0 \over \lambda}\right)^2\,,
\end{equation}
i.e., sinuosity \mpar{R3.3}\new{$\xi$ is  related to the undulation parameter $w_0/\lambda$.} 

On the other hand, using \eqref{eq:w2}, \eqref{eq:dreff} and \eqref{eq:sinuosity-small}, 
in the long-time limit $L_d(t)\gg \lambda$ we obtain
\begin{equation} \label{eq:reff-w0}
r_\text{eff}^2-\langle r^2\rangle_v \simeq w_0^2\,.
\end{equation}
Here we assumed that the undulation is small ($w_0\ll \lambda$, which is the case based on our estimates using realistic axons, \figref{fig:appendix-undulation-simulation}c), and to the first order in the small parameter $w_0/\lambda$ we approximated $z\simeq l$, and $\cos kz\simeq\cos kl$. 
The double integral $\int\! \dots dl dl'$ is then equivalent to averaging 
$$
\left< (\cos kl - \cos kl')^2 \right> 
= 4\left< \sin^2 {k(l+l')\over 2} \sin^2 {k(l-l')\over 2} \right>_{l,l'} = 1
$$
approximately independently over $l+l'$ and $l-l'$. 

Eliminating the amplitude $w_0$ using \eqref{eq:reff-w0} yields
\begin{equation} \label{eq:dreff-sinuosity}
\xi-1 \simeq \left(\frac{\pi}{\lambda}\right)^2 \cdot \left(r_\text{eff}^2-\langle r^2\rangle_v\right)\,,
\end{equation}
which indicates that, by plotting each axon's \new{$\xi-1$ versus its $r_\text{eff}^2-\langle r^2\rangle_v$, the slope $=(\pi/\lambda)^2$} provides an estimate of the undulation wavelength; thus we find \new{$\lambda\approx 26$ $\mu$m} based on \figref{fig:IAS-result}g.

\subsection{Diffusion time-dependence of the axonal undulation}
The RD contributed by the axonal undulation is quantified as $\langle \delta w^2\rangle/4t$ in \figref{fig:appendix-undulation-simulation}a. At short times, $\langle \delta w^2\rangle/4t$ is the projection of the along-skeleton Gaussian diffusion onto the direction transverse to the main direction, as demonstrated in \figref{fig:appendix-undulation-simulation}b:
\begin{equation} \label{eq:D-undulation-short-time}
\left.\frac{\langle \delta w^2\rangle}{4t}\right|_{t\to0} = D_a\cdot\frac{\langle \sin^2\theta\rangle}{2}\,,
\end{equation}
where $\theta = \theta(z)$ is the angle between the individual axon's skeleton segment at $z$ and its main direction. The factor 2 in denominator is because in realistic IAS, the axonal undulation happens in two directions (e.g., $x$- and $y$-axes) 
perpendicular to the main direction, along which the ``macroscopic curvature'' is ignored. At long times, $\langle \delta w^2\rangle$ approaches  a constant (\figref{fig:IAS-result}e), and $\langle \delta w^2\rangle/4t$ scales as $1/t$ (\figref{fig:appendix-undulation-simulation}a).

To better understand the diffusivity time-dependence due to the axonal undulation, we decompose an undulating axon via multiple harmonics with amplitude ($w_{xn}$, $w_{yn}$), wave number ($k_n$) and phase ($\phi_{xn}$, $\phi_{yn}$) for the $n$-th harmonic ($n\in \mathbb{N}$):
\begin{equation} \label{eq:multi-harmonic}
{\bf w} = \sum_n w_{xn} \cos(k_n z + \phi_{xn}) {\bf \hat{x}} + w_{yn} \cos(k_n z + \phi_{yn}) {\bf \hat{y}}\,.
\end{equation}
Here, we focus on the case of $k_n=n\cdot 2\pi/L_z$, such that all harmonics are orthogonal to each other. 
\new{The exact relation between $z$ and $l$ becomes complicated, and here we only provide the lowest-order approximate solution. Substituting \eqref{eq:multi-harmonic} into the Taylor expansion of \eqref{eq:sinuosity-exact} for $|d{\bf w}/dz|\ll 1$ and setting integration bounds from 0 to $z(l)$, we obtain
\begin{equation} \label{eq:l-z}
    l \simeq \xi \cdot z\,,\quad w_{0n}\ll 1/k_n\,,
\end{equation}
where
\begin{equation} \notag
    \xi \simeq 1+\sum_n \sum_{m=1}^\infty \binom{\frac{1}{2}}{m} \binom{2m}{m}\cdot \left(\frac{w_{0n}^2 k_n^2}{4}\right)^m = 1+\sum_n \frac{1}{4}w_{0n}^2k_n^2+\dots\,,
\end{equation}
is the approximated sinuosity, and $w_{0n}\equiv\sqrt{w_{xn}^2+w_{yn}^2}$ is the 2{\it d} undulation amplitude. In realistic IAS, $w_{0n}\lesssim$ 0.6 $\mu$m for $n = 1$, $2$, $3$ (\figref{fig:appendix-undulation-simulation}c).}
Assuming that the undulation is so small ($w_{0n}\ll1/k_n$) that $z\simeq l/\xi$, we can approximate $\cos(k_n z+\phi_n)\simeq \cos(k_n l/\xi+\phi_n)$. Substituting into \eqref{eq:w2}, 
integrating  over $l+l'$ and $l-l'$, 
and using for each harmonic 
$\langle \sin^2 [k_n(l+l')/2 + \phi_n] \rangle_{l+l'} = 1/2$ and 
$$
\int\! {dz\over \sqrt{4\pi D_\infty^\parallel}} \sin^2 {k_n z\over 2}\, e^{-z^2/4D_\infty^\parallel t} = \frac12 \left( 1-e^{-D_\infty^\parallel k_n^2 t} \right), 
$$
we obtain
\begin{equation} \label{eq:w2-multi-harmonics}
\langle \delta w^2\rangle \simeq \sum_n w_{0n}^2 
\left( 1-e^{- D_\infty^\parallel k_n^2 t}\right)\,, \quad w_{0n}\ll1/k_n,
\end{equation}
\new{with $D_\infty^\parallel = D_a/\xi^2$. In \ref{sec:app-undulation-axial}, we will prove that $D_\infty^\parallel$ is the bulk diffusivity along the axon in $t\to\infty$ limit, whereas in the main text, we approximated $D_\infty^\parallel\simeq D_a$ since $\xi\gtrsim 1$ for small undulations ($w_{0n}k_n\ll 1$).}

At short times, $\langle \delta w^2\rangle \simeq \sum_n w_{0n}^2 k_n^2 D_\infty^\parallel t$. Substituting into \eqref{eq:D-undulation-short-time}, the projection factor transverse to the main direction is given by
\begin{equation*}
\langle \sin^2\theta \rangle \simeq \frac{2}{\xi^2} \sum_n \left(\frac{\pi w_{0n}}{\lambda_n} \right)^2\,,
\end{equation*}
where $\lambda_n = 2\pi/k_n$ is the corresponding wavelength. At long times, $\langle \delta w^2\rangle \simeq \sum_n w_{0n}^2$, leading to a diffusivity $\langle \delta w^2\rangle/4t$ scaling as $1/t$.

To further discuss the diffusion time-dependence due to the axonal undulation, we calculated the instantaneous diffusivity \citep{novikov2014meso} \new{in $d=2$-dimensions} based on \eqref{eq:w2-multi-harmonics}:
\begin{equation} \label{eq:Dinst-undulation}
\frac{1}{2\new{d}}\partial_t\langle \delta w^2\rangle\simeq \sum_n \frac{1}{4}\cdot\frac{w_{0n}^2}{t_n} \cdot e^{-t/t_n}\,,\quad w_{0n} \ll 1/k_n \,,
\end{equation}
where $t_n=1/D_\infty^\parallel k_n^2 \propto \lambda_n^2/D_\infty^\parallel$ is the correlation time related to the undulation wavelength. Based on the realistic axonal skeleton, the instantaneous diffusivity due to the axonal undulation does not decrease mono-exponentially over diffusion times (\figref{fig:appendix-undulation-simulation}d), indicating that the axonal undulation consists of more than one harmonic. 

\new{
Note that for each harmonic, the instantaneous diffusivity is given by an exponentially decaying function, which has the same functional form as the contribution of each Laplace eigenmode in a fully confined pore. This is not surprising, since the motion projected on the plane transverse to the axon axis is confined. 

In a standard way, the instantaneous diffusivity in \eqref{eq:Dinst-undulation} due to undulations can be translated into a frequency-dependent dispersive diffusivity \citep{burcaw2015meso,novikov2019note}:
\begin{subequations}
\begin{align} \label{eq:D-undul-disp-a}
    {\cal D}_u(\omega) &= -i\omega\int_0^\infty dt \, e^{i\omega t} \frac{1}{2d}\partial_t\langle \delta w^2\rangle\\ \label{eq:D-omega-b}
    &\simeq \sum_n \frac{1}{4} \cdot \frac{w_{0n}^2}{t_n} \cdot\frac{-i\omega t_n}{1-i\omega t_n}\\
    &= \sum_n \frac{1}{4} \cdot \frac{w_{0n}^2}{t_n} \cdot \left( -i\omega t_n + \omega^2 t_n^2 + ... \right)\,,
\end{align}
\end{subequations}
which yields the oscillating-gradient measured $\omega$-dependent diffusivity due to undulations in the limit of large number of oscillations \citep{novikov2019note}:
\begin{equation} \label{eq:D-undul-ogse}
    \text{Re } {\cal D}_u(\omega) \simeq \left( \sum_n \frac{1}{4} w_{0n}^2 t_n \right)\cdot \omega^2 + {\cal O}(\omega^4).
\end{equation}
This $\omega^2$-dependence at low frequencies is expected due to the nature of confined diffusion \citep{burcaw2015meso} of axonal undulations projected to the $x$-$y$ plane (\figref{fig:appendix-undulation-shape}c).

The knowledge of ${\cal D}_u(\omega)$ in \eqref{eq:D-omega-b} enables ones to evaluate the effect of any sequence on the second order cumulant of the diffusion signals \citep{novikov2011ogse}:
\begin{equation} \label{eq:gpa}
    -\ln S = \int \frac{d\omega}{2\pi}\,{\cal D}_u(\omega) |q_\omega|^2 + {\cal O}(g^4)\,,
\end{equation}
where $q_\omega$ is the Fourier transform of the diffusion wave vector $q(t)=\int_0^t\, dt'\, g(t')$. For a pulsed-gradient sequence of diffusion time $t$ and pulse width $\delta$, we have \citep{callaghan1991book,burcaw2015meso}
\begin{equation} \label{eq:q-omega-pg}
    q_\omega = \frac{g}{(i\omega)^2} \left(e^{i\omega \delta}-1\right) \left(e^{i\omega t}-1\right)\,.
\end{equation}
Substituting Eqs.~(\ref{eq:D-omega-b}) and (\ref{eq:q-omega-pg}) into \eqref{eq:gpa}, we obtain the pulsed-gradient measured radial diffusivity $\left.-\tfrac{1}{b}\ln S\right|_{b\to0}$ due to undulations:
\begin{multline} \label{eq:D-undul-wide}
    D_u(t,\delta)
    \simeq \sum_n \frac{1}{4}\cdot \frac{w_{0n}^2 t_n^2}{\delta^2(t-\delta/3)} \cdot \left[ 2\frac{\delta}{t_n} -2 \right.\\ \left.+ 2e^{-t/t_n} + 2e^{-\delta/t_n} - e^{-(t-\delta)/t_n} - e^{-(t+\delta)/t_n} \right]\,,
\end{multline}
which is analogous to the solution of axon caliber in Eq. [11] of \citep{vangelderen1994wide}. In other words, it is difficult to distinguish contributions of axon caliber and axonal undulation at the level of $g^2$.
The frequency integral is calculated along the real axis on the complex plane of $\omega$ with two poles at $-i/ t_n$ and $-i0$ on the lower-half plane of $\omega$ \citep{novikov2010emt}.

In the wide pulse limit of undulations, i.e., $\delta\gg t_n\sim \lambda_n^2/D_\infty^\parallel$, the RD due to undulations acquires the \citet{neuman1974gpa} form
\begin{equation} \label{eq:D-undul-neuman}
    D_u(t,\delta)\simeq \sum_n \frac{1}{8\pi^2}\cdot\frac{w_{0n}^2\lambda_n^2}{D_\infty^\parallel\delta}\cdot\frac{1}{t-\delta/3}\,, \quad \delta\gg t_n\,,
\end{equation}
whose functional form is the same as that in \eqref{eq:D-neuman}, although the corresponding undulation length scale is not of the same order as the caliber length scale in realistic axonal shapes: In our mouse brain EM sample of CC, the length scale of caliber variations is $\langle r^4\rangle_v\sim(0.78$ $\mu$m$)^4$ in \eqref{eq:D-neuman}, whereas the length scale of undulations is $w_{0n}\lesssim0.6$ $\mu$m (\figref{fig:appendix-undulation-simulation}d) and $\lambda_n\sim26$ $\mu$m (\eqref{eq:dreff-sinuosity} and \figref{fig:IAS-result}g) in \eqref{eq:D-undul-neuman}, i.e., $\langle r^4\rangle_v\ll w_{0n}^2\lambda_n^2$. Therefore, in the wide pulse limit, the axon size estimation based on the RD time-dependence is dominated by undulations if the undulation effects are not partly factored out by spherically averaging signals.
}


\setcounter{figure}{0}
\section{Bias in the axon size caused by equal-step-length random leap}
\label{sec:app-radius}
In our simulations, the interaction of random walkers and diffusion boundaries (e.g., cell membrane, myelin sheath) is modeled as the equal-step-length random leap: a step crossing the boundary is canceled, and another direction will be chosen to leap until the step does not cross any boundaries. This strategy largely reduces the calculation time of MC simulations.

However, walkers are effectively repelled away from the boundary since the step crossing boundaries is rejected by the algorithm. To calculate the theoretical values of $\langle r^2\rangle_v$, \new{$\langle r^4\rangle_v$} and $K_\infty$ based on Eqs.~(\ref{eq:reff}), (\ref{eq:K}) and (\ref{r62}), the cylindrical radius $r$ needs to be corrected to match the simulation results in \figref{fig:bead-result}. The corrected radius $r'$ is smaller than the actual radius $r$ of the geometry due to this repulsion effect. \new{In fact, the first order correction $\delta r\equiv r-r'$ of effective radius solely depends on the step size $\delta s$, and not on the cylinder radius $r$ (details to be published elsewhere): 
\begin{equation} \label{eq:dr-3d}
    \delta r = 
    \begin{cases} 
        \delta s/4 & d=1\,,\\
        \delta s/(2\pi) & d=2\,,\\
        \delta s/8 & d=3\,.
    \end{cases}
\end{equation}

}

\setcounter{figure}{0}
\new{
\section{Axonal undulation and diffusion along axons}
\label{sec:app-undulation-axial}

For completeness, it is worthwhile to explore the diffusivity time-dependence {\it along} the  undulating axons. 
By setting $\bf{\hat{n}}={\bf\hat{z}}$ in \eqref{eq:l-n-2}, the second-order cumulant $\langle \delta z^2\rangle$ of the diffusion displacement projected onto the main axis is given by
\begin{equation} \label{eq:z2}
    \langle \delta z^2\rangle \simeq \frac{1}{C} \int \left|z(l)-z(l')\right|^2\cdot e^{-\frac{(l-l')^2}{4D_a t}}\,dl \,dl'\,,
\end{equation}
where $C$ is the same normalization factor as in \eqref{eq:w2}, and $D_a$ is diffusivity along the axonal skeleton. In the narrow pulse limit, the (apparent) axial diffusivity due to undulation is by definition $D^\parallel(t) \equiv \langle \delta z^2\rangle/(2t)$. 
Substituting \eqref{eq:l-z} into \eqref{eq:z2}, we obtain the bulk diffusivity along axons: 
\begin{align} \notag
    D^\parallel(t) &\simeq \frac{D_a}{\xi^2} \equiv D_\infty^\parallel\,, \quad t\to\infty \\ \label{eq:D-ax-und}
    &\simeq D_a\left(1-\sum_n\frac{1}{2}w_{0n}^2k_n^2\right)\,.
\end{align}

To explore the axial diffusivity time-dependence, it is necessary to derive  $z=z(l)$ entering \eqref{eq:z2}. 
For simplicity, here we provide an approximate solution of the 1-harmonic model, perturbatively in the undulation amplitude $w_0$. The expansion is controlled by the small parameter 
$\epsilon \equiv (w_0 k/2)^2 = (\pi w_0/\lambda)^2 \ll 1$. To the first order in $\epsilon$, the sinuosity (\ref{eq:sinuosiy-1-harmonic}) is $\xi = 1+\epsilon$. In what follows, we will need to explore $z=z(l)$ up to $\epsilon^2$, since the time-dependence will turn out to be an ${\cal O}(w_0^4)$ effect. 

Expanding the definition (\ref{eq:sinuosity-exact}) for the 1-harmonic model (\ref{eq:1-harmonic}) up to $\epsilon^4$, we find
\begin{eqnarray*}
l(z) &\simeq& \int_0^z\! dz \left(1+ \frac{(dw/dz)^2}{2} - \frac{(dw/dz)^4}{8} \right) \\
&=& \xi z - (\epsilon-\epsilon^2)\, \frac{\sin(2kz)}{2k} - \epsilon^2\, \frac{\sin(4kz)}{16k} \,,
\end{eqnarray*}
where $\xi(\epsilon)= 1+\epsilon-\frac{3}{4}\epsilon^2 + {\cal O}(\epsilon^3)$ is defined after \eqref{eq:l-z}.
We then invert the above relation perturbatively to obtain $z=z(l)$. 
After some trigonometry needed to keep the term $\sin(2kz) \approx \sin (2kl/\xi) + (\epsilon/2) \sin (4kl) $ 
exact up to ${\cal O}(\epsilon)$, and substituting $\sin 4kz \to \sin 4kl$ as that term is already ${\cal O}(\epsilon^2)$, we obtain 
\begin{equation} \notag
    z(l)\simeq \frac{l}{\xi} \cdot\left[1+\frac{\epsilon}{2kl}\cdot\sin\left(\frac{2kl}{\xi}\right) 
    - \frac{\epsilon^2}{2kl} \left(\sin (2kl) -\frac{5}{8} \sin (4kl) \right) \right]\,.
\end{equation}
Substituting $z(l)$ into \eqref{eq:z2}, we expand the square $\left(z(l)-z(l')\right)^2$ up to $\epsilon^2$.  
We see that the first term in the above expression for $z(l)$, $(l-l')^2/\xi^2$, yields $D_\infty$. 
For the other terms employing the trigonometric functions, we change variables to $l_+ = (l+l')/2$ and $l_- = l-l'$, 
and observe that most of them vanish after integration over $dl_+$ in the limit of axon length $L\to \infty$, including the hard-earned $\epsilon^2$ terms from the third term in the above $z(l)$. 
The only nonzero term at the $\epsilon^2$ level comes from the second term in $z(l)$ above, 
$\sim \epsilon^2 (\sin(2kl)-\sin(2kl'))^2 \sim \epsilon^2 \cos^2 (2kl_+) \sin^2 (kl_-)$. 
Remarkably, this means that $l(z)$ expanded up to ${\cal O}(\epsilon)$ only would be enough --- 
but it was not obvious from the outset. Using the integration result right before Eq.~(\ref{eq:w2-multi-harmonics}), 
we finally obtain
\begin{align} \notag
    D^\parallel(t) = \frac{\langle \delta z^2\rangle}{2t} &\simeq D_\infty^\parallel + c^\parallel\cdot\frac{1}{t}\left(1-e^{-4 D_\infty^\parallel k^2 t}\right)\,,\\ \label{eq:ad-undulation}
    &\simeq D_\infty^\parallel + \frac{c^\parallel}{t}\,,\quad t\gg\frac{1}{4 D_\infty^\parallel k^2}\,,
\end{align}
where
\begin{equation} \notag
    D_\infty^\parallel = \frac{D_a}{\xi^2}\,,\quad c^\parallel = \frac{\epsilon^2}{8k^2}\,,
\end{equation}
or equivalently, by dropping additional higher order terms of $\epsilon$, we have
\begin{equation} \label{eq:D-c-1h-axial}
    \frac{D_a-D_\infty^\parallel}{D_\infty^\parallel}\simeq \frac{1}{2}w_0^2k^2\sim \frac{w_0^2}{\lambda^2}\,,\quad c^\parallel \simeq \frac{1}{128}w_0^4k^2 \sim \frac{w_0^4}{\lambda^2}\,.
\end{equation}
In other words, at fixed $w_0$, the longer the wavelength, the closer the $D_\infty^\parallel$ is to the free $D_a$, and the smaller the $c^\parallel$. 

\begin{figure*}[tb!]\centering
\includegraphics[width=0.75\textwidth]{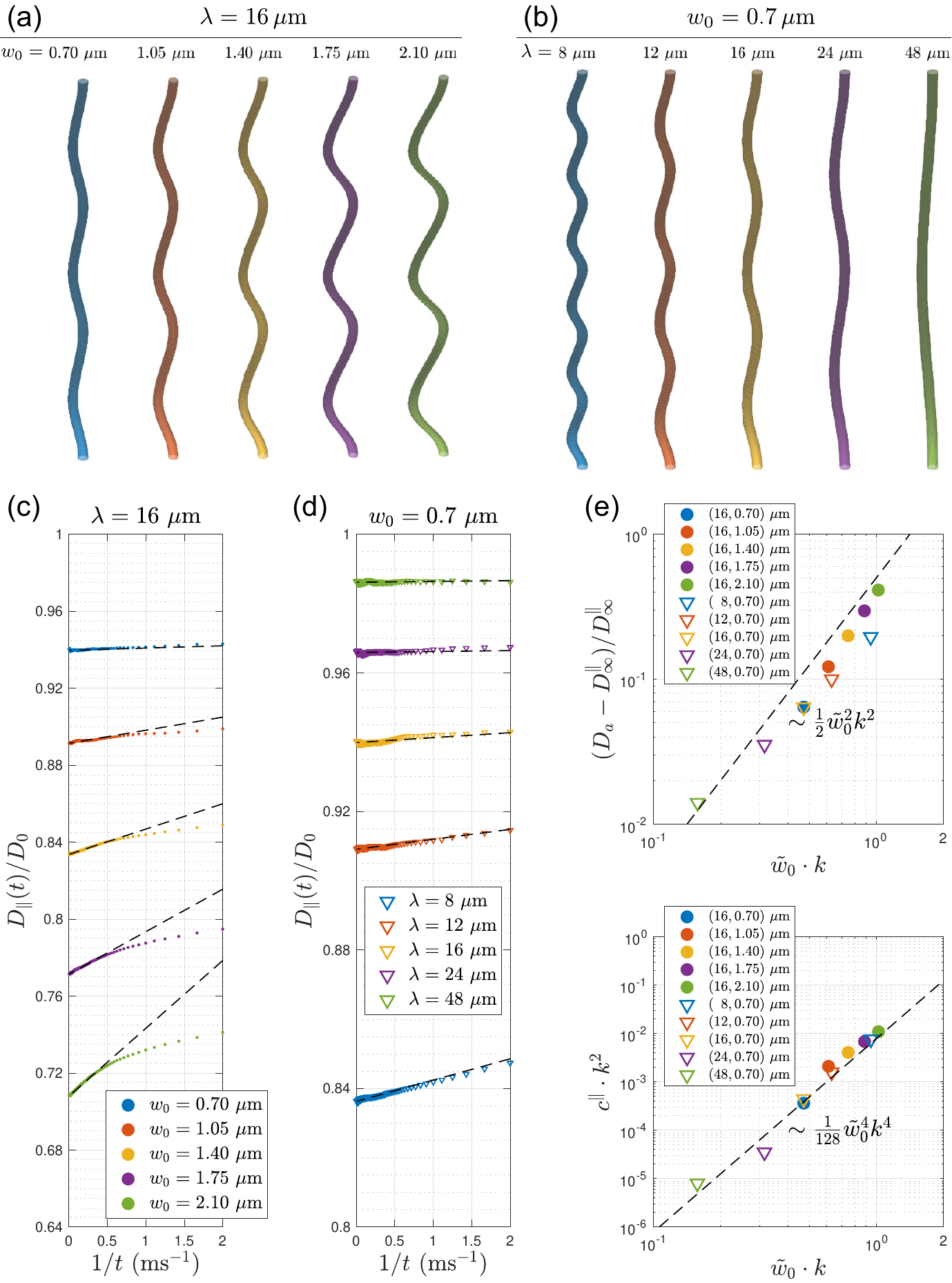}
\caption{Simulated axial diffusivites $D_\parallel(t)$ along undulating axons with a constant radius $r=0.5$ $\mu$m. The undulation of each axon is composed of only one harmonic with (a) undulation wavelength $\lambda=16$ $\mu$m and undulation amplitude $w_0=0.70$, $1.05$, $1.40$, $1.75$, or $2.10$ $\mu$m, respectively, and (b) undulation amplitude $w_0=0.7$ $\mu$m and undulation wavelength $\lambda=8$, $12$, $16$, $24$, and $48$ $\mu$m, respectively. (c-d) The simulated $D_\parallel(t)$ time-dependence scales as $1/t$ at long $t$, as predicted in \eqref{eq:ad-undulation}. (e) The fit parameters $D_\infty^\parallel$ and $c^\parallel$ in \eqref{eq:ad-undulation} are consistent with the theoretical prediction in \eqref{eq:D-c-1h-axial}, after the shift $w_0\to\tilde{w}_0\simeq w_0+r$. Notably, the above simulations demonstrate that the $1/t$-dependence in axial diffusivity is applicable to 1-harmonic undulating fibers in a wide range of $\tilde{\epsilon}\equiv (\tilde{w}_0k/2)^2$.  }
\label{fig:appendix-undulation-axial}
\end{figure*}

To validate the $1/t$-dependence of axial diffusivity due to undulations in \eqref{eq:ad-undulation} and \eqref{eq:D-c-1h-axial}, we performed MC simulations in undulating axons with a constant radius $r=0.5$ $\mu$m along axons. Each axon in the first setting has a 1-harmonic undulation with the wavelength $\lambda=16$ $\mu$m and amplitudes $w_0=0.7$, $1.05$, $1.40$, $1.75$, and $2.10$ $\mu$m, respectively (\figref{fig:appendix-undulation-axial}a), and each axon in the second setting has a 1-harmonic undulation of the amplitude $w_0=0.7$ $\mu$m and wavelengths $\lambda=2\pi/k=8$, $12$, $16$, $24$, and $48$ $\mu$m, respectively (\figref{fig:appendix-undulation-axial}b).
The above fibers yield $\epsilon\simeq0.002-0.17<1$, where taking only the first few terms of $\xi(\epsilon)$ is a good approximation.
For each run, $2\times10^7$ random walkers diffuse within each axon with the same parameters settings as in the narrow pulse regime in the main text, except for the implementation of elastic collisions. The simulated axial diffusivity $D_\parallel(t)$ has a tiny $1/t$-dependence ($<0.2\%$ and $<0.07\%$ diffusivity change at $t=20-100$ ms in \figref{fig:appendix-undulation-axial}c and d respectively), as predicted in \eqref{eq:ad-undulation}.

While the 1-harmonic model assumed zero axon radius, in our MC simulations the radius was of the order of the undulation amplitude. It turns out that the effect of the finite radius was not negligible. 
Empirically, based on our MC simulations (\figref{fig:appendix-undulation-axial}e), we find that the ``interference" of the undulation amplitude $w_0$ and axon radius $r$ is best dealt with by shifting $w_0\to \tilde{w}_0\simeq w_0+r$ in \eqref{eq:D-c-1h-axial} (one can rationalize that finite radius makes the maximal deviation from axon axis more pronounced by about $\sim r$). The above fibers for simulations (\figref{fig:appendix-undulation-axial}a-b) yield $\epsilon\to \tilde{\epsilon}\equiv(\tilde{w}_0 k/2)^2 \simeq 0.006-0.26<1$, where taking the first few terms of $\xi(\epsilon)\to\xi(\tilde{\epsilon})$ is still a reasonable approximation.

We also note that the characteristic $1/t$-dependence of axial diffusivity in \eqref{eq:ad-undulation} (and the exponential one for the corresponding $D_{\rm inst}(t))$ is caused by the 1{\it d} {\it periodic} restrictions of the 1-harmonic undulation along the axon. For an undulating axon composed of {\it randomnly placed undulations} (e.g., harmonics with randomly varying phases),  one expects a diffusivity $1/\sqrt{t}$-dependence along the axon \citep{novikov2014meso}.

}



\begin{thebibliography}{60}
\expandafter\ifx\csname natexlab\endcsname\relax\def\natexlab#1{#1}\fi
\expandafter\ifx\csname url\endcsname\relax
  \def\url#1{\texttt{#1}}\fi
\expandafter\ifx\csname urlprefix\endcsname\relax\def\urlprefix{URL }\fi

\bibitem[{Abdollahzadeh et~al.(2019)Abdollahzadeh, Belevich, Jokitalo, Tohka,
  and Sierra}]{abdollahzadeh20193dem}
Abdollahzadeh, A., Belevich, I., Jokitalo, E., Tohka, J., Sierra, A., 2019.
  Automated 3d axonal morphometry of white matter. Scientific Reports 9~(1),
  6084.

\bibitem[{Alexander et~al.(2010)Alexander, Hubbard, Hall, Moore, Ptito, Parker,
  and Dyrby}]{alexander2010activeax}
Alexander, D.~C., Hubbard, P.~L., Hall, M.~G., Moore, E.~A., Ptito, M., Parker,
  G.~J., Dyrby, T.~B., 2010. Orientationally invariant indices of axon diameter
  and density from diffusion mri. NeuroImage 52~(4), 1374--1389.

\bibitem[{Assaf et~al.(2008)Assaf, Blumenfeld-Katzir, Yovel, and
  Basser}]{assaf2008axcaliber}
Assaf, Y., Blumenfeld-Katzir, T., Yovel, Y., Basser, P.~J., 2008. Axcaliber: a
  method for measuring axon diameter distribution from diffusion mri. Magnetic
  Resonance in Medicine 59~(6), 1347--1354.

\bibitem[{Bain and Meaney(2000)}]{bain2000undulation}
Bain, A.~C., Meaney, D.~F., 2000. Tissue-level thresholds for axonal damage in
  an experimental model of central nervous system white matter injury. Journal
  of Biomechanical Engineering 122~(6), 615--622.

\bibitem[{Bain et~al.(2003)Bain, Shreiber, and Meaney}]{bain2003undulation}
Bain, A.~C., Shreiber, D.~I., Meaney, D.~F., 2003. Modeling of microstructural
  kinematics during simple elongation of central nervous system tissue. Journal
  of Biomechanical Engineering 125~(6), 798--804.

\bibitem[{Barazany et~al.(2009)Barazany, Basser, and Assaf}]{Barazany2009}
Barazany, D., Basser, P.~J., Assaf, Y., 2009. {In vivo measurement of axon
  diameter distribution in the corpus callosum of rat brain}. Brain 132,
  1210--1220.

\bibitem[{Baron et~al.(2015)Baron, Kate, Gioia, Butcher, Emery, Budde, and
  Beaulieu}]{baron2015stroke}
Baron, C.~A., Kate, M., Gioia, L., Butcher, K., Emery, D., Budde, M., Beaulieu,
  C., 2015. Reduction of diffusion-weighted imaging contrast of acute ischemic
  stroke at short diffusion times. Stroke 46~(8), 2136--2141.

\bibitem[{Benjamini et~al.(2016)Benjamini, Komlosh, Holtzclaw, Nevo, and
  Basser}]{benjamini2016wmadm}
Benjamini, D., Komlosh, M.~E., Holtzclaw, L.~A., Nevo, U., Basser, P.~J., 2016.
  White matter microstructure from nonparametric axon diameter distribution
  mapping. NeuroImage 135, 333--344.

\bibitem[{Bouchaud and Georges(1990)}]{bouchaud1990review}
Bouchaud, J.-P., Georges, A., 1990. Anomalous diffusion in disordered media:
  statistical mechanisms, models and physical applications. Physics Reports
  195~(4-5), 127--293.

\bibitem[{Brabec et~al.(2020)Brabec, Lasi{\v{c}}, and
  Nilsson}]{brabec2019undulation}
Brabec, J., Lasi{\v{c}}, S., Nilsson, M., 2020. Time-dependent diffusion in
  undulating thin fibers: Impact on axon diameter estimation. NMR in
  Biomedicine 33, e4187.

\bibitem[{Budde and Frank(2010)}]{budde2010bead}
Budde, M.~D., Frank, J.~A., 2010. Neurite beading is sufficient to decrease the
  apparent diffusion coefficient after ischemic stroke. Proceedings of the
  National Academy of Sciences 107~(32), 14472--14477.

\bibitem[{Burcaw et~al.(2015)Burcaw, Fieremans, and Novikov}]{burcaw2015meso}
Burcaw, L.~M., Fieremans, E., Novikov, D.~S., 2015. Mesoscopic structure of
  neuronal tracts from time-dependent diffusion. NeuroImage 114, 18--37.

\bibitem[{Callaghan(1991)}]{callaghan1991book}
Callaghan, P.~T., 1991. Principles of nuclear magnetic resonance microscopy.
  Clarendon Press.

\bibitem[{Callaghan et~al.(1991)Callaghan, Coy, MacGowan, Packer, and
  Zelaya}]{callaghan1991pore}
Callaghan, P.~T., Coy, A., MacGowan, D., Packer, K.~J., Zelaya, F.~O., 1991.
  Diffraction-like effects in nmr diffusion studies of fluids in porous solids.
  Nature 351~(6326), 467.

\bibitem[{Dhital et~al.(2018)Dhital, Kellner, Kiselev, and
  Reisert}]{dhital2018restricted}
Dhital, B., Kellner, E., Kiselev, V.~G., Reisert, M., 2018. The absence of
  restricted water pool in brain white matter. NeuroImage 182, 398--406.

\bibitem[{Dhital et~al.(2019)Dhital, Reisert, Kellner, and
  Kiselev}]{dhital2019ias}
Dhital, B., Reisert, M., Kellner, E., Kiselev, V.~G., 2019. Intra-axonal
  diffusivity in brain white matter. NeuroImage 189, 543--550.

\bibitem[{Duval et~al.(2015)Duval, McNab, Setsompop, Witzel, Schneider, Huang,
  Keil, Klawiter, Wald, and Cohen-Adad}]{duval2015spinal}
Duval, T., McNab, J.~A., Setsompop, K., Witzel, T., Schneider, T., Huang,
  S.~Y., Keil, B., Klawiter, E.~C., Wald, L.~L., Cohen-Adad, J., 2015. In vivo
  mapping of human spinal cord microstructure at 300 m{T}/m. NeuroImage 118,
  494--507.

\bibitem[{Fieremans et~al.(2016)Fieremans, Burcaw, Lee, Lemberskiy, Veraart,
  and Novikov}]{fieremans2016time}
Fieremans, E., Burcaw, L.~M., Lee, H.-H., Lemberskiy, G., Veraart, J., Novikov,
  D.~S., 2016. In vivo observation and biophysical interpretation of
  time-dependent diffusion in human white matter. NeuroImage 129, 414--427.

\bibitem[{Fieremans et~al.(2011)Fieremans, Jensen, and
  Helpern}]{fieremans2011dki}
Fieremans, E., Jensen, J.~H., Helpern, J.~A., 2011. White matter
  characterization with diffusional kurtosis imaging. NeuroImage 58~(1),
  177--188.

\bibitem[{Fieremans and Lee(2018)}]{fieremans2018phantom}
Fieremans, E., Lee, H.-H., 2018. Physical and numerical phantoms for the
  validation of brain microstructural mri: A cookbook. NeuroImage 182, 39--61.

\bibitem[{Fieremans et~al.(2010)Fieremans, Novikov, Jensen, and
  Helpern}]{fieremans2010karger}
Fieremans, E., Novikov, D.~S., Jensen, J.~H., Helpern, J.~A., 2010. Monte carlo
  study of a two-compartment exchange model of diffusion. NMR in Biomedicine
  23~(7), 711--724.

\bibitem[{Fontana(1781)}]{fontana1781undulation}
Fontana, F., 1781. Trait{\'e} sur le v{\'e}nin de la vip{\`e}re, sur les
  poisons am{\'e}ricaines. Vol.~2. chez Nyon l'Ain{\'e}.

\bibitem[{Garthwaite et~al.(1999)Garthwaite, Brown, Batchelor, Goodwin, and
  Garthwaite}]{garthwaite1999ischemic}
Garthwaite, G., Brown, G., Batchelor, A., Goodwin, D., Garthwaite, J., 1999.
  Mechanisms of ischaemic damage to central white matter axons: a quantitative
  histological analysis using rat optic nerve. Neuroscience 94~(4), 1219--1230.

\bibitem[{Jacobs(1935)}]{jacobs1935fjeq}
Jacobs, M.~H., 1935. Diffusion processes. In: Diffusion Processes. Springer,
  pp. 1--145.

\bibitem[{Jensen and Helpern(2010)}]{jensen2010dki}
Jensen, J.~H., Helpern, J.~A., 2010. {MRI} quantification of non-gaussian water
  diffusion by kurtosis analysis. NMR in Biomedicine 23~(7), 698--710.

\bibitem[{Jensen et~al.(2005)Jensen, Helpern, Ramani, Lu, and
  Kaczynski}]{jensen2005dki}
Jensen, J.~H., Helpern, J.~A., Ramani, A., Lu, H., Kaczynski, K., 2005.
  Diffusional kurtosis imaging: the quantification of non-gaussian water
  diffusion by means of magnetic resonance imaging. Magnetic Resonance in
  Medicine 53~(6), 1432--1440.

\bibitem[{Jespersen et~al.(2013)Jespersen, Lundell, S{\o}nderby, and
  Dyrby}]{jespersen2013pfg}
Jespersen, S.~N., Lundell, H., S{\o}nderby, C.~K., Dyrby, T.~B., 2013.
  Orientationally invariant metrics of apparent compartment eccentricity from
  double pulsed field gradient diffusion experiments. NMR in Biomedicine
  26~(12), 1647--1662.

\bibitem[{Johnson et~al.(2013)Johnson, Stewart, and Smith}]{johnson2013tbi}
Johnson, V.~E., Stewart, W., Smith, D.~H., 2013. Axonal pathology in traumatic
  brain injury. Experimental Neurology 246, 35--43.

\bibitem[{Kaden et~al.(2016)Kaden, Kruggel, and Alexander}]{kaden2016smt}
Kaden, E., Kruggel, F., Alexander, D.~C., 2016. Quantitative mapping of the
  per-axon diffusion coefficients in brain white matter. Magnetic Resonance in
  Medicine 75~(4), 1752--1763.

\bibitem[{Kiselev and Novikov(2018)}]{Kiselev2018}
Kiselev, V.~G., Novikov, D.~S., 2018. {Transverse NMR relaxation in biological
  tissues}. NeuroImage 182, 149--168.

\bibitem[{Lasi{\v{c}} et~al.(2014)Lasi{\v{c}}, Szczepankiewicz, Eriksson,
  Nilsson, and Topgaard}]{lasic2014magicangle}
Lasi{\v{c}}, S., Szczepankiewicz, F., Eriksson, S., Nilsson, M., Topgaard, D.,
  2014. Microanisotropy imaging: quantification of microscopic diffusion
  anisotropy and orientational order parameter by diffusion mri with
  magic-angle spinning of the q-vector. Frontiers in Physics 2, 11.

\bibitem[{Lee et~al.(2018)Lee, Fieremans, and Novikov}]{lee2018rd}
Lee, H.-H., Fieremans, E., Novikov, D.~S., 2018. What dominates the time
  dependence of diffusion transverse to axons: Intra-or extra-axonal water?
  NeuroImage 182, 500--510.

\bibitem[{Lee et~al.(2020)Lee, Papaioannou, Kim, Novikov, and
  Fieremans}]{lee2019axial}
Lee, H.-H., Papaioannou, A., Kim, S.-L., Novikov, D.~S., Fieremans, E., 2020. A
  time-dependent diffusion {MRI} signature of axon caliber variations and
  beading. Communications Biology 3, 354.

\bibitem[{Lee et~al.(2019)Lee, Yaros, Veraart, Pathan, Liang, Kim, Novikov, and
  Fieremans}]{lee20193dem}
Lee, H.-H., Yaros, K., Veraart, J., Pathan, J.~L., Liang, F.-X., Kim, S.~G.,
  Novikov, D.~S., Fieremans, E., 2019. Along-axon diameter variation and axonal
  orientation dispersion revealed with 3{D} electron microscopy: implications
  for quantifying brain white matter microstructure with histology and
  diffusion {MRI}. Brain Structure and Function, 1--20.

\bibitem[{Li and Murphy(2008)}]{li2008reperfusion}
Li, P., Murphy, T.~H., 2008. Two-photon imaging during prolonged middle
  cerebral artery occlusion in mice reveals recovery of dendritic structure
  after reperfusion. Journal of Neuroscience 28~(46), 11970--11979.

\bibitem[{McKinnon et~al.(2017)McKinnon, Jensen, Glenn, and
  Helpern}]{mckinnon2017highb}
McKinnon, E.~T., Jensen, J.~H., Glenn, G.~R., Helpern, J.~A., 2017. Dependence
  on b-value of the direction-averaged diffusion-weighted imaging signal in
  brain. Magnetic Resonance Imaging 36, 121--127.

\bibitem[{Mitra et~al.(1992)Mitra, Sen, Schwartz, and
  Le~Doussal}]{mitra1992propagator}
Mitra, P.~P., Sen, P.~N., Schwartz, L.~M., Le~Doussal, P., 1992. Diffusion
  propagator as a probe of the structure of porous media. Physical Review
  Letters 68~(24), 3555.

\bibitem[{Neuman(1974)}]{neuman1974gpa}
Neuman, C., 1974. Spin echo of spins diffusing in a bounded medium. The Journal
  of Chemical Physics 60~(11), 4508--4511.

\bibitem[{Nilsson et~al.(2012)Nilsson, L{\"a}tt, St{\aa}hlberg, van Westen, and
  Hagsl{\"a}tt}]{nilsson2012undulation}
Nilsson, M., L{\"a}tt, J., St{\aa}hlberg, F., van Westen, D., Hagsl{\"a}tt, H.,
  2012. The importance of axonal undulation in diffusion {MR} measurements: a
  {Monte Carlo} simulation study. NMR in Biomedicine 25~(5), 795--805.

\bibitem[{Novikov et~al.(2019)Novikov, Fieremans, Jespersen, and
  Kiselev}]{novikov2019note}
Novikov, D.~S., Fieremans, E., Jespersen, S.~N., Kiselev, V.~G., 2019.
  Quantifying brain microstructure with diffusion {MRI}: Theory and parameter
  estimation. NMR in Biomedicine 32~(4), e3998.

\bibitem[{Novikov et~al.(2014)Novikov, Jensen, Helpern, and
  Fieremans}]{novikov2014meso}
Novikov, D.~S., Jensen, J.~H., Helpern, J.~A., Fieremans, E., 2014. Revealing
  mesoscopic structural universality with diffusion. Proceedings of the
  National Academy of Sciences 111~(14), 5088--5093.

\bibitem[{Novikov and Kiselev(2010)}]{novikov2010emt}
Novikov, D.~S., Kiselev, V.~G., 2010. Effective medium theory of a
  diffusion-weighted signal. NMR in Biomedicine 23~(7), 682--697.

\bibitem[{Novikov and Kiselev(2011)}]{novikov2011ogse}
Novikov, D.~S., Kiselev, V.~G., 2011. Surface-to-volume ratio with oscillating
  gradients. Journal of Magnetic Resonance 210~(1), 141--145.

\bibitem[{Novikov et~al.(2018)Novikov, Veraart, Jelescu, and
  Fieremans}]{novikov2018rotinv}
Novikov, D.~S., Veraart, J., Jelescu, I.~O., Fieremans, E., 2018.
  Rotationally-invariant mapping of scalar and orientational metrics of
  neuronal microstructure with diffusion {MRI}. NeuroImage 174, 518--538.

\bibitem[{Palombo et~al.(2018)Palombo, Ligneul, Hernandez-Garzon, and
  Valette}]{palombo2018metabolite}
Palombo, M., Ligneul, C., Hernandez-Garzon, E., Valette, J., 2018. Can we
  detect the effect of spines and leaflets on the diffusion of brain
  intracellular metabolites? NeuroImage 182, 283--293.

\bibitem[{Palombo et~al.(2016)Palombo, Ligneul, Najac, Le~Douce, Flament,
  Escartin, Hantraye, Brouillet, Bonvento, and Valette}]{palombo2016metabolite}
Palombo, M., Ligneul, C., Najac, C., Le~Douce, J., Flament, J., Escartin, C.,
  Hantraye, P., Brouillet, E., Bonvento, G., Valette, J., 2016. New paradigm to
  assess brain cell morphology by diffusion-weighted {MR} spectroscopy in vivo.
  Proceedings of the National Academy of Sciences 113~(24), 6671--6676.

\bibitem[{Ronen et~al.(2014)Ronen, Budde, Ercan, Annese, Techawiboonwong, and
  Webb}]{ronen2014cc}
Ronen, I., Budde, M., Ercan, E., Annese, J., Techawiboonwong, A., Webb, A.,
  2014. Microstructural organization of axons in the human corpus callosum
  quantified by diffusion-weighted magnetic resonance spectroscopy of
  n-acetylaspartate and post-mortem histology. Brain Structure and Function
  219~(5), 1773--1785.

\bibitem[{Schilling et~al.(2016)Schilling, Janve, Gao, Stepniewska, Landman,
  and Anderson}]{schilling20163dconfocal}
Schilling, K., Janve, V., Gao, Y., Stepniewska, I., Landman, B.~A., Anderson,
  A.~W., 2016. Comparison of 3{D} orientation distribution functions measured
  with confocal microscopy and diffusion {MRI}. NeuroImage 129, 185--197.

\bibitem[{Schilling et~al.(2018)Schilling, Janve, Gao, Stepniewska, Landman,
  and Anderson}]{schilling2018histmri}
Schilling, K.~G., Janve, V., Gao, Y., Stepniewska, I., Landman, B.~A.,
  Anderson, A.~W., 2018. Histological validation of diffusion {MRI} fiber
  orientation distributions and dispersion. NeuroImage 165, 200--221.

\bibitem[{Sepehrband et~al.(2016)Sepehrband, Alexander, Kurniawan, Reutens, and
  Yang}]{sepehrband2016adm}
Sepehrband, F., Alexander, D.~C., Kurniawan, N.~D., Reutens, D.~C., Yang, Z.,
  2016. Towards higher sensitivity and stability of axon diameter estimation
  with diffusion-weighted {MRI}. NMR in Biomedicine 29~(3), 293--308.

\bibitem[{Shepherd and Raastad(2003)}]{shepherd2003varicosity}
Shepherd, G.~M., Raastad, M., 2003. Axonal varicosity distributions along
  parallel fibers: a new angle on a cerebellar circuit. The Cerebellum 2~(2),
  110--113.

\bibitem[{Shepherd et~al.(2002)Shepherd, Raastad, and
  Andersen}]{shepherd2002varicosity}
Shepherd, G.~M., Raastad, M., Andersen, P., 2002. General and variable features
  of varicosity spacing along unmyelinated axons in the hippocampus and
  cerebellum. Proceedings of the National Academy of Sciences 99~(9),
  6340--6345.

\bibitem[{Szczepankiewicz et~al.(2015)Szczepankiewicz, Lasi{\v{c}}, van Westen,
  Sundgren, Englund, Westin, St{\aa}hlberg, L{\"a}tt, Topgaard, and
  Nilsson}]{szczepankiewicz2015magicangle}
Szczepankiewicz, F., Lasi{\v{c}}, S., van Westen, D., Sundgren, P.~C., Englund,
  E., Westin, C.-F., St{\aa}hlberg, F., L{\"a}tt, J., Topgaard, D., Nilsson,
  M., 2015. Quantification of microscopic diffusion anisotropy disentangles
  effects of orientation dispersion from microstructure: applications in
  healthy volunteers and in brain tumors. NeuroImage 104, 241--252.

\bibitem[{Tang-Schomer et~al.(2012)Tang-Schomer, Johnson, Baas, Stewart, and
  Smith}]{tang2012microtubule}
Tang-Schomer, M.~D., Johnson, V.~E., Baas, P.~W., Stewart, W., Smith, D.~H.,
  2012. Partial interruption of axonal transport due to microtubule breakage
  accounts for the formation of periodic varicosities after traumatic axonal
  injury. Experimental Neurology 233~(1), 364--372.

\bibitem[{Vangelderen et~al.(1994)Vangelderen, DesPres, Vanzijl, and
  Moonen}]{vangelderen1994wide}
Vangelderen, P., DesPres, D., Vanzijl, P., Moonen, C., 1994. Evaluation of
  restricted diffusion in cylinders. phosphocreatine in rabbit leg muscle.
  Journal of Magnetic Resonance, Series B 103~(3), 255--260.

\bibitem[{Veraart et~al.(2019)Veraart, Fieremans, and
  Novikov}]{veraart2019highb}
Veraart, J., Fieremans, E., Novikov, D.~S., 2019. On the scaling behavior of
  water diffusion in human brain white matter. NeuroImage 185, 379--387.

\bibitem[{Veraart et~al.(2020)Veraart, Nunes, Rudrapatna, Fieremans, Jones,
  Novikov, and Shemesh}]{veraart2020highb}
Veraart, J., Nunes, D., Rudrapatna, U., Fieremans, E., Jones, D.~K., Novikov,
  D.~S., Shemesh, N., 2020. Noninvasive quantification of axon radii using
  diffusion {MRI}. eLife 9.

\bibitem[{West et~al.(2016)West, Kelm, Carson, and Does}]{west2016gratio}
West, K.~L., Kelm, N.~D., Carson, R.~P., Does, M.~D., 2016. A revised model for
  estimating g-ratio from {MRI}. NeuroImage 125, 1155--1158.

\bibitem[{Xing et~al.(2013)Xing, Lin, Wu, and Gong}]{xing2013randomleap}
Xing, H., Lin, F., Wu, Q., Gong, Q., 2013. Investigation of different boundary
  treatment methods in {Monte-Carlo} simulations of diffusion {NMR}. Magnetic
  Resonance in Medicine 70~(4), 1167--1172.

\bibitem[{Zimmerman and Brittin(1957)}]{zimmerman1957R2}
Zimmerman, J., Brittin, W.~E., 1957. Nuclear magnetic resonance studies in
  multiple phase systems: lifetime of a water molecule in an adsorbing phase on
  silica gel. The Journal of Physical Chemistry 61~(10), 1328--1333.

\end{thebibliography}

\restoregeometry
\end{document}